\documentclass[useAMS,usenatbib]{mn2e}
\usepackage{float,graphics,epsfig,array,amsmath,amssymb}
\usepackage{times}

\title
[{\em Suzaku} X-ray spectrum of NGC\,4051]
{
Spectral variability and reverberation time delays in the {\em Suzaku} X-ray spectrum of NGC\,4051.
}

\author
[L.~Miller et al.]
{L.~Miller$^{1}$,
\thanks{E-mail: L.Miller@physics.ox.ac.uk},
T.~J.~Turner$^{2,3}$,
J.~N.~Reeves$^{4}$,
A.~Lobban$^{4}$,
S.~B.~Kraemer$^{5,3}$
and 
D.~M.~Crenshaw$^{6}$.\\
$^{1}$Dept. of Physics, Oxford University, 
Denys Wilkinson Building, Keble Road, Oxford OX1 3RH, U.K.\\
$^{2}$Dept. of Physics, University of Maryland Baltimore County, Baltimore, MD 21250, U.S.A.\\
$^{3}$Astrophysics Science Division, NASA/GSFC, Greenbelt, MD 20771, U.S.A.\\
$^{4}$Astrophysics Group, School of Physical and Geographical Sciences, 
Keele University, Keele, Staffordshire ST5 8EH, U.K.\\
$^{5}$Institute for Astrophysics and Computational Sciences, Department of Physics, 
The Catholic University of America, Washington, DC 20064, U.S.A.\\
$^{6}$Department of Physics and Astronomy, Georgia State University,
Astronomy Offices, One Park Place South SE, Suite 700, Atlanta, GA 30303, U.S.A.
}

\begin{document}

\date{Accepted 2009 December 1. Received 2009 December 1; in original form 2009 September 13}

\pagerange{\pageref{firstpage}--\pageref{lastpage}} \pubyear{2009}

\maketitle

\label{firstpage}

\begin{abstract}
Long-exposure {\em Suzaku} X-ray observations of the nearby active galaxy NGC\,4051 from 2005 and 2008
are analysed, in an attempt to reach a self-consistent understanding of both the
spectral variability on long timescales and the broad-band variability at high time 
resolution.  The techniques of principal components analysis and a maximum likelihood
method of power spectrum analysis are used.  In common with other type\,I 
active galactic nuclei (AGN), the spectral
variability is dominated by a varying-normalisation power-law component
together with a quasi-steady, hard-spectrum offset
component that contains Fe\,K atomic features.  NGC\,4051 displays a strong excess over a power-law
at energies above
20\,keV, some fraction of which also appears to vary with the power-law continuum.
The fast timescale power spectrum has a shape generally consistent
with previous determinations, with the previously-known dependence
on broad-band photon energy,
but in the new data significant differences are found between
the low and high flux states of the source, demonstrating the power spectrum is non-stationary.
Frequency-dependent time lags between the hard and soft bands 
of up to $970\pm 225$\,s are measured.  
The existence of the observed time lags excludes the possibility that the hard spectral component
originates as reflection from the inner accretion disk.  
We instead show that the time lags and their frequency- and energy-dependence may be explained simply
by the effects of reverberation in the hard band, caused by reflection from a thick shell
of material with maximum lags of about
10\,000\,s.  If the reflecting material surrounds the AGN, it extends to a distance about
$1.5 \times 10^{14}$\,cm, 600 gravitational radii, from the illuminating source and the
global covering factor is $C_g \ga 0.44$, confirming previous suggestions that type\,I AGN
have high covering factors of absorbing and reflecting material.
Given the spectral and timing similarities with other
type\,I AGN, we infer that this source structure is common in the type\,I population.
\end{abstract}


\section{Introduction}

Given the impossibility of imaging the innermost regions of active galactic nuclei (AGN) 
at X-ray energies, the principal tools for studying accretion in this waveband
have been either modelling of
X-ray spectra or studying the rapid X-ray variability of nearby, bright AGN.  Owing to
signal-to-noise limitations spectral analysis has tended to concentrate on fitting simple
composite models to mean spectra \citep[e.g.][]{miniutti07a}.
For the same reason, timing analyses have studied the
variability on short timescales but in broad bandpasses of photon energy \citep[e.g.][]{mchardy04a}.
To date, conclusions from these studies have not been combined in a self-consistent way.  
However, there is substantially more information available in the spectral variability than
is contained in the mean spectrum alone, and now
that some long-exposure observations have been made, those
data may be time-sliced and the spectral variability used as a powerful constraint on the models
(e.g. \citealt{miller07a}, \citealt*{miller08a}).  Long exposure observations have also revealed
interesting variability signatures on short timescales (see below).
In this paper we aim to carry out both spectral variability
and timing analysis on long-duration {\em Suzaku} observations of the nearby AGN NGC\,4051, and
to reach a consistent understanding of the source behaviour.

\begin{table*}
\caption{Summary of the three {\em Suzaku} observations, giving the 
ID number, dates, the detectors used in the analysis, 
total duration, {\sc xis} on-source time, the
mean {\sc xis} count rate in the 0.3-10\,keV range, and the mean flux
in the 2-10\,keV range.
The 2005
count rate has been scaled to the value for two detectors, for consistency with
the 2008 observations.
\label{table:data}}
\begin{tabular}{lllcccc}
ID & dates & detectors & duration & on-time & mean count rate & mean 2-10\,keV flux \\
   &       &           &   /s     &   /s    &      /s$^{-1}$  & /10$^{-11}$\,erg\,s$^{-1}$cm$^{-2}$\\
\hline
700004010 & 2005 Nov 10-13 & {\sc xis}\,0,2,3, {\sc pin} & 222592 & 119618 & 0.93 & 0.87 \\
703023010 & 2008 Nov 6-12  & {\sc xis}\,0,3, {\sc pin} & 494834 & 274531 & 4.19 & 2.41 \\
703023020 & 2008 Nov 23-25 & {\sc xis}\,0,3, {\sc pin} & 161695 & \phantom{0}78394 & 2.87 & 1.79 \\
\hline
\end{tabular}
\end{table*}

There are two phenomena of particular interest in each of the ``spectral variability'' 
and the ``timing'' domains.  
The first is that it is now well-established that type\,I AGN frequently show
a broad ``red wing'' of emission extending below the 6.4\,keV\,Fe\,K$\alpha$ emission line
that appears as an excess above a model power-law continuum \citep[e.g.][]{tanaka95a}, and this
as been widely interpreted as being Fe\,K$\alpha$ emission from within a few Schwarzschild radii
of the black hole.  Model fits to mean spectra infer a redshift so large that the emission
would need to come from within the innermost stable circular orbit of a Schwarzschild black
hole, leading to claims of the detection of black hole 
spin in these AGN \citep[e.g.][]{wilms01a}.  However, a particular
problem is that any Fe\,K$\alpha$ emission should vary in phase with the illuminating continuum,
whereas the red wing appears to be largely invariant while the illuminating source appears to
exhibit large time variations \citep[e.g.][]{iwasawa96a, vaughan04a}.  The problem is now known to
extend to high energies: above 20\,keV many AGN have an excess of continuum compared with
extrapolation of a 2-10\,keV power-law model, and this has been supposed to be continuum reflection
(Compton scattering modified by the photoelectric opacity of the gas) 
from the same material that produces the red wing.  Indeed, this ``hard excess''
is likewise known to exhibit little variability compared with the 2-10\,keV band, pointing to
a common origin with the red wing \citep{miniutti07a}. However, its flux is so large that it cannot
readily be ascribed to reflection unless the illuminating source is significantly obscured along
our line of sight \citep[e.g.][]{terashima09a}.

To overcome these problems, 
\citet{fabian03a}, \citet{miniutti03a} and \citet{miniutti04a} have proposed a ``light-bending'' model
in which the effect of light following geodesics near the black hole mean that a
reflection component from the inner accretion disk 
may be made relatively invariant, if it is illuminated by a source whose position changes with respect
to the disk.  The model requires both that the illuminating
source is compact, comparable in size to the event horizon, and moves towards and away from the
disk.  As it moves, the flux received by a distant observer varies as light is removed from
the line of sight. In this model, it is the change in position of the primary source 
that dominates the observed variability, rather than intrinsic fluctuations.  The light bending
could also explain the high apparent reflectivity in the hard band.
However, \citet{miller08a} and \citet*{miller09a} have shown that the red wing and the
hard excess in MCG--6-30-15 may instead be explained by the spectrum expected from clumpy, 
``partial-covering''
absorbers, and that a model in which the absorber covering fraction varies provides a good description of
the observed spectral variability, as well as of the mean spectrum of this AGN.  Both Compton-scattering
reflection and absorption shape this hard spectral component, but the relative importance of 
scattering/reflection and absorbed transmission is not yet established \citep{miller09a}.

The second important phenomenon occurs in the timing domain.  There are now a significant number of
AGN where time lags in the range 10s to 1000s of seconds between different spectral bands are
seen, in the sense that hard-band photons lag soft-band 
(\citealt*{papadakis01a}, \citealt*{vaughan03a}, \citealt{mchardy04a}, \citealt{markowitz05a},
\citealt{arevalo06a}, \citealt{markowitz07a}, \citealt{arevalo08a}).
Importantly, the lags are dependent on the frequency of
the source variation, in the sense that if a source's variations are decomposed into Fourier modes,
the lag between hard and soft energy bands
increases with the period of those modes, and also increases with the separation in energy of
the bands.  The lags and their frequency-dependence have been explained by a model 
comprising perturbations propagating inwards on an accretion disk where harder X-ray emission
is emitted from smaller radii \citep{arevalo06b}.
Although the initial observations were
consistent with a linear relationship with Fourier period, recently, more complex behaviour has
been seen, leading to a hypothesis that fluctuations
arise in multiple components \citep{mchardy07a}.  However, this explanation of rapid
fluctuations propagating through the accretion disk is not consistent with the light-bending
model, which has the primary variations caused by motion of the hypothesised compact source
above the accretion disk.
It is our aim in this paper to reconcile the phenomena of spectral variability and 
time lags in a detailed study of NGC\,4051.

NGC\,4051 is a well-studied, nearby narrow-line Type\,I AGN with redshift $z=0.0023$.  
If the redshift were 
purely due to cosmological Hubble flow
its distance would be 9.3\,Mpc for Hubble's constant $H_0=74$\,km\,s$^{-1}$\,Mpc$^{-1}$,
but for such a nearby galaxy the Tully-Fisher distance is probably more reliable,
placing it at 15.2\,Mpc \citep{russell04a}.  It has a black hole mass determined
from optical reverberation mapping of 
$M_{BH} = 1.7^{+0.55}_{-0.52} \times 10^6$\,M$_{\odot}$ \citep{denney09a}.  
It has long been known to be highly variable on short timescales at X-ray energies,
with spectral variability that is correlated with flux 
(\citealt{lawrence85a, lawrence87a}, \citealt{matsuoka90a}, \citealt{kunieda92a},
\citealt{mchardy95a}, \citealt{lamer03a}).
The timing properties of X-ray observations with the
{\em Rossi X-ray Timing Explorer} (RXTE) and {\em XMM-Newton} have been studied in
detail by \citet{mchardy04a}.  The X-ray spectrum has previously been studied by
\citet{guainazzi96a, guainazzi98a} and
\citet{pounds04c}, among others.  Of particular interest is a 2005
{\em Suzaku} observation obtained when the source was in a low state 
that reveals a strong excess of emission above a power-law at energies $E \ga 20$\,keV
\citep{terashima09a}.
Overall, the consensus from recent analyses is that NGC\,4051 has low states where the primary power-law
source has largely disappeared and where the remaining emission appears to be dominated
by reflected emission \citep{guainazzi98a, pounds04c, terashima09a}.  
\citet{terashima09a} found that partially-covered reflection
was required to fit the hard excess, rather than supposing this to be inner-disk 
gravitationally-redshifted reflection.  The high
variability on short timescales means that this AGN provides a powerful testbed for using
spectral variability and power spectrum analysis methods
to untangle the source emission components.

We now consider both the 2005 {\em Suzaku} observation and two new {\em Suzaku}
observations obtained in 2008, and analyse both the time-dependent spectrum 
and the high time resolution behaviour. 

\section{Observations and data reduction}
\label{sec:data}
The {\em Suzaku} data analysed here were reduced as described by 
\citet{turner09c} and Lobban et~al. (in preparation).
The {\em Suzaku} X-ray Imaging Spectrometer \citep[{{\sc xis}},][]{koyama07}
comprises four X-ray telescopes \citep{mitsuda07} each
with a CCD detector.  {\sc xis} CCDs 0, 2, 3 are configured to be
front-illuminated and yield useful data over $\sim 0.6-10.0$\,keV
with energy resolution FWHM\,$\sim 150\,$eV at 6\,keV. {\sc xis}\,1 is a
back-illuminated CCD and has an enhanced soft-band response down to
0.2\,keV but lower area at 6\,keV than the front-illuminated CCDs, as well as a higher
background level at high energies. {\sc xis}\,1 was not used in our analysis.
Use of {\sc xis}\,2 was discontinued after a charge leak was discovered
in Nov 2006, so we used data from {\sc xis}\,0, 2 and 3 in 2005,
and data from {\sc xis}\,0 and 3 in 2008.

Our analysis used three observations 
from 2005 Nov 10-13, 2008 Nov 6-12 and 2008 Nov 23-25. 
The data were reduced using v6.4.1 of {\sc HEAsoft} and screened to
 exclude periods, i) during and within 500 seconds of the South Atlantic
 Anomaly, ii) with an Earth elevation angle less than 10$^\circ$ and
 iii) with cut-off rigidity $>6$ GeV. The source was observed at the nominal
 pointing for the {\sc xis}. The front-illuminated CCDs were in $3 \times 3$ and $5
 \times 5$ edit-modes, with normal clocking mode.  For the {\sc xis} we
 selected events with grades 0,2,3,4, and 6 and removed hot and
 flickering pixels using the {\sc sisclean} script.  The spaced-row charge
 injection was used.  The {\sc xis} products were extracted from
 circular regions of 2.9\arcmin radius with background spectra from
 a region of the same size, offset from the source (avoiding the
 calibration sources at the edges of the chips). The response and
 ancillary response files were created using {\sc xisrmfgen v2007 May}
 and {\sc xissimarfgen v2008 Mar}.

{\em Suzaku} also has a non-imaging, collimated Hard X-ray Detector \citep[HXD,][]{takahashi07}
whose {\sc pin} instrument provides useful data over 15-70\,keV for bright AGN.
NGC\,4051 is too faint to be detected by the HXD GSO instrument, but
was detected in the {\sc pin}. For the analysis we used the model ``D''
background \citep{fukazawa09a}.
As the {\sc pin} background rate is strongly variable around
the orbit, we first selected source data to discard events within 
500s of a South Atlantic Anomaly passage, we also rejected events with day/night elevation
angles $> 5^\circ$.  The time filter resulting from the screening
was then applied to the background events model file to give {\sc pin}
model background data for the same time intervals covered by the
on-source data. As the background events file was generated using ten
times the actual background count rate, an adjustment to the
background spectrum was applied to account for this factor.  
{\sc hxddtcor v2007 May} was run to apply the deadtime correction to the
source spectrum. 
As the {\sc pin} deadtime correction varies, 
for the time-dependent analysis below, the 
correction was calculated at 128\,s time resolution and the mean
correction in each analysed time interval was applied to the source counts.

The observations are summarised in Table\,\ref{table:data}, which also gives
the {\sc xis} count rates and the flux integrated over 2-10\,keV.  The 2005 observation
caught NGC\,4051 in a historical low state, while in 2008 the source flux
was more typical, and close to the value in the 2001 {\em XMM-Newton} data
analysed by \citet{mchardy04a}.

\begin{figure}
\resizebox{0.45\textwidth}{!}{
\rotatebox{-90}{
\includegraphics{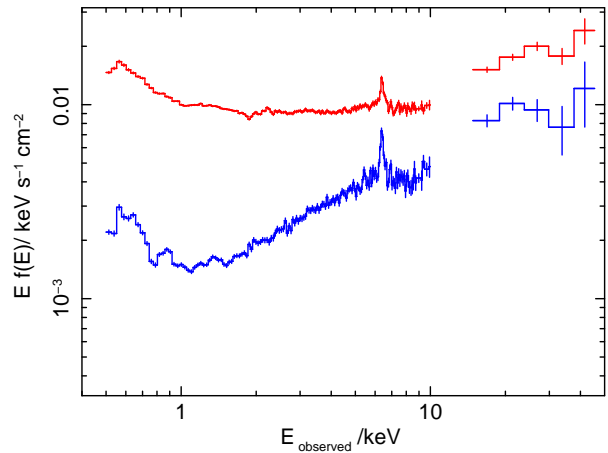}
}}
\caption{
Combined {\sc xis} and {\sc pin} spectra of NGC\,4051 from 2005 (lower curve)
and 2008 Nov\,6-12 (upper curve), plotted as $Ef(E)$, unfolded against a power-law
with photon index $\Gamma=2$.
{\sc xis} data have been binned at the instrumental HWHM, {\sc pin} data at intervals
$\Delta\log_{10}E=0.1$.
Photon shot noise errors are shown.  No cross-calibration correction between {\sc xis}
and {\sc pin} instruments has been applied in this figure.
}
\label{fig:spectra}
\end{figure}

\begin{figure*}
\resizebox{0.45\textwidth}{!}{
\rotatebox{-90}{
\includegraphics{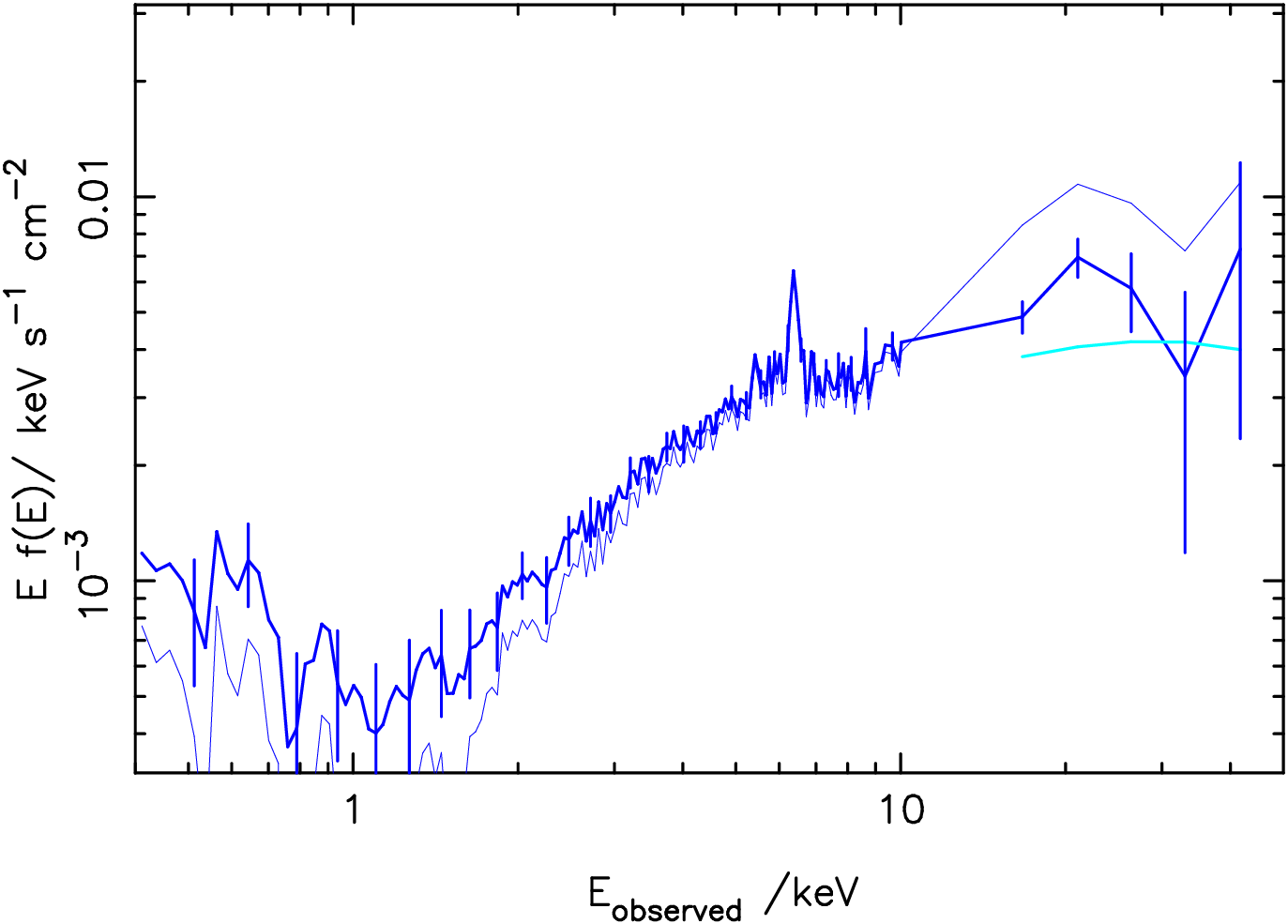}
}}
\hspace*{5mm}
\resizebox{0.45\textwidth}{!}{
\rotatebox{-90}{
\includegraphics{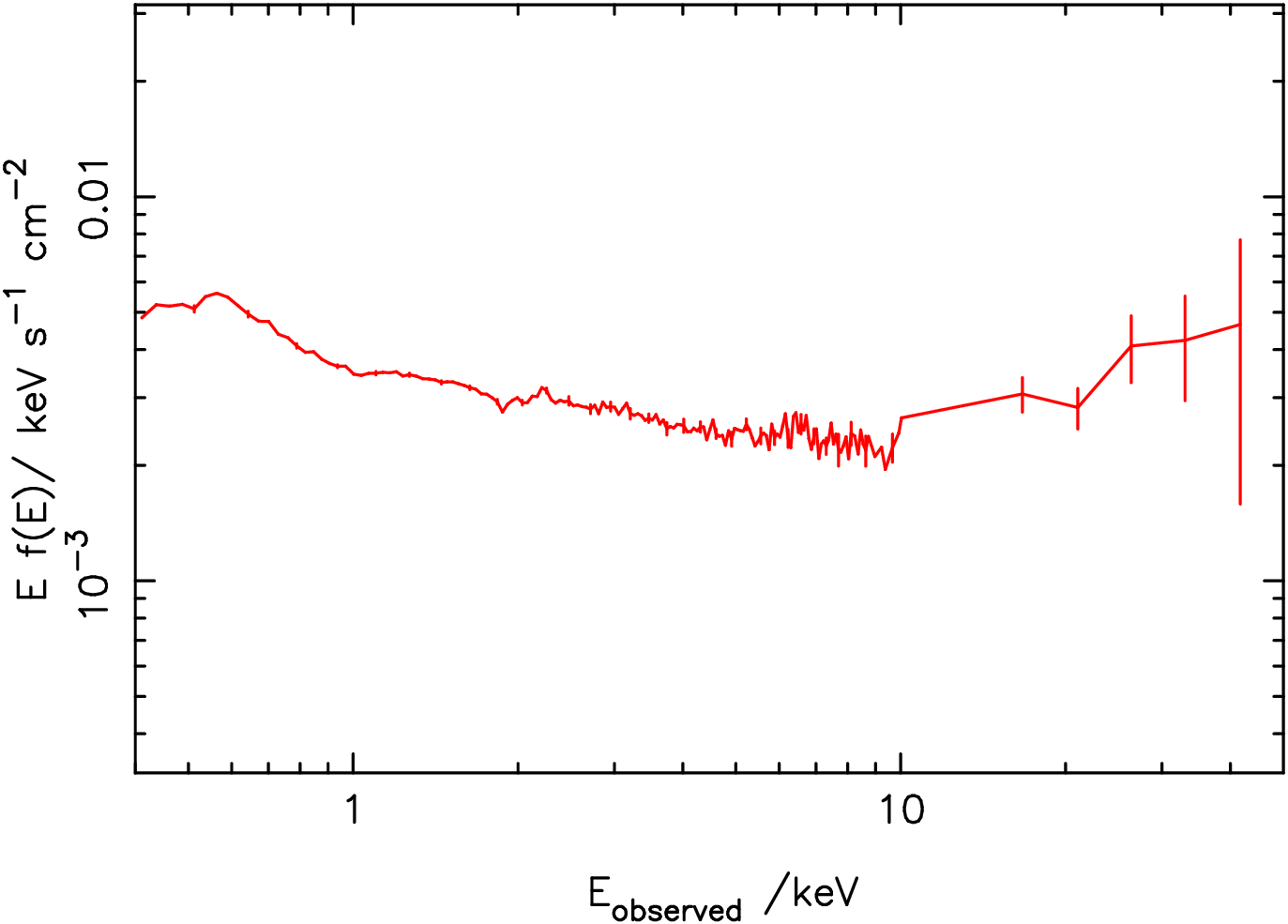}
}}
\caption{
Principal components analysis of the combined 2005-8 dataset for NGC\,4051: (left)
the offset component, showing the possible range of this component (upper and lower
curves) and the component of cosmic X-ray background that has been subtracted (solid curve).
For clarity, errors are shown only every fifth point on one of the offset spectra;
(right) eigenvector one.
}
\label{fig:pca}
\end{figure*}

\section{Spectral variability on timescales longer than 20\,ks}
\label{sec:pca}
\subsection{Principal components analysis}
Our first step to understanding the variations in NGC\,4051 is to carry out a 
principal components analysis
(PCA) of the spectral variations.  This is a method of decomposing the variations
into orthogonal components, eigenvectors.  By ordering the eigenvectors
according to their eigenvalues we can extract the principal modes of variation.
It is particularly well suited to variations that are additive in nature
(such as variations in partial covering fraction by a clumpy absorber).
We use the method described in detail by \citet{miller07a}, in which spectral
bins are created of width equal to half the energy-dependent instrument
resolution.  Spectra are averaged over and sampled at time intervals that are
sufficiently long that variations between time samplings are dominated by intrinsic
variations rather than shot noise.  The method uses
singular value decomposition to cope with the problem that there are
more spectral points than there are available time bins: in this case 
the leading eigenvectors are still defined and may be extracted.

In interpreting the results, there should be no expectation that the orthogonal
eigenspectra produced correspond uniquely to any physical component in the
AGN.  However, what has been found in previous analyses of other AGN \citep{miller07a, miller08a}
is that only a small number of variable components are required to describe the
spectral variability, which is indicative that additive processes are at work.
In those previous analyses we concentrated on eigenvector one, which describes
the primary source variation, and subsumed the rest of the source spectrum into
an ``offset'' component which is essentially steady in time (see \citealt{miller07a}
and \citealt{turner07a} for more detailed discussion).  The offset component
is not uniquely defined, because we are free to add or subtract arbitrary amounts 
of eigenvector one without changing the decomposition.  However, we may define
a minimum offset component by requiring that no spectral bin have negative flux
and a maximum offset component by requiring that no spectral bin be higher than
the lowest observed value.

\subsection{Spectral variability}
\label{sec:spectralvariability}
Fig.\,\ref{fig:spectra} shows the mean spectra from 2005 and 2008 Nov\,6-12, plotted as unfolded
spectra against a power-law with photon index $\Gamma=2$.
In the figure, a cosmic background model has been subtracted from the {\sc pin} data by using
{\sc xspec} v\,11.3.2ag to generate a
spectrum from a cosmic X-ray background model \citep{gruber99} normalized to the $34^\prime
\times 34^\prime$ {\sc pin} field of view, which was then combined with the {\sc pin}
instrument background file.  
No correction for the calibration difference between
the {\sc xis} and {\sc pin} instruments has been applied in Fig.\,\ref{fig:spectra}: this
has the effect of making the {\sc pin} data too high by 16\,percent relative to the {\sc xis} data.
The basic spectral variability mode of the source is clearly seen, with the hard, low-state
and higher softer spectrum at the two epochs.  There is a significant hard excess above
20\,keV, which has varied between the epochs.  In 2008 there is also a ``red wing'' as
an excess above a power-law continuum below the 6.4\,keV\,Fe\,K$\alpha$ line: in the 2005
data this can be seen to have developed into a rather different continuum shape.
Previous analyses have assumed the red wing to be relativistically broadened Fe\,K$\alpha$
emission \citep[e.g.][]{lamer03a} but this interpretation is not obviously supported by the
PCA.

The PCA of the observations in the energy range 0.4-50\,keV is shown in Fig.\,\ref{fig:pca}.
The time sampling was 20\,ks: although this source is highly variable on shorter
timescales, it is important to avoid the regime where counts in each spectral bin
are dominated by shot noise.  The full 2005-8 dataset was used, incorporating both
{\sc xis} and {\sc pin} data.  {\sc xis} data were binned at the spectral resolution
half-width at half-maximum.
To achieve adequate signal-to-noise in the {\sc pin} data, coarser spectral
binning was used ($\Delta\log_{10}E = 0.1$). 
The minimum and maximum limits on the offset component
are shown, together with the contribution of the cosmic X-ray background that has
been subtracted from the {\sc pin} offset component.  
The {\sc pin} data have been reduced by the 16\,percent cross-calibration factor
between the {\sc xis} and {\sc pin} instruments.
Uncertainties on each spectral point were obtained by a 
Monte Carlo method as described by \citet{miller07a}.
Eigenvector one accounts for 81\,percent of the variance in this energy band.  Eigenvector
two accounts for a further 11\,percent of the variance, and although noisy this component
appears dominated by a residual anticorrelation between the {\sc pin} and {\sc xis}
data, which when considered with the eigenvector one spectrum
implies there is a minor element of uncorrelated variation between the {\sc xis} and {\sc pin}
bands.  In this paper we shall concentrate on the gross spectral variability characterised
by eigenvector one and the offset component.

\begin{figure*}
\resizebox{0.95\textwidth}{!}{
\rotatebox{-90}{\includegraphics{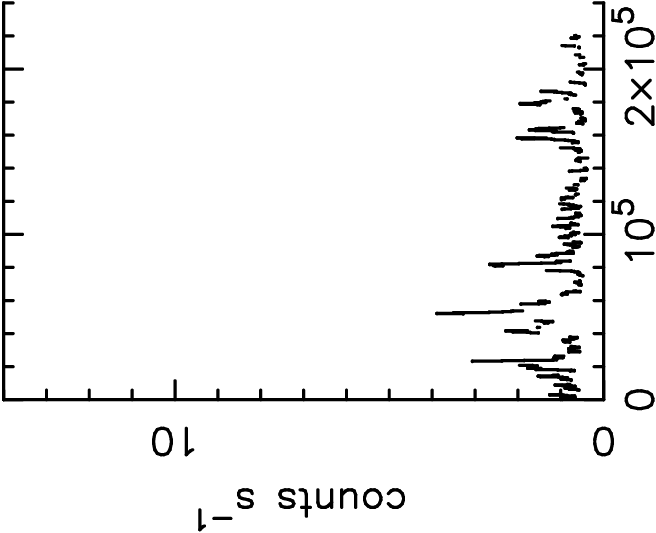}}
\hspace*{3mm}
\rotatebox{-90}{\includegraphics{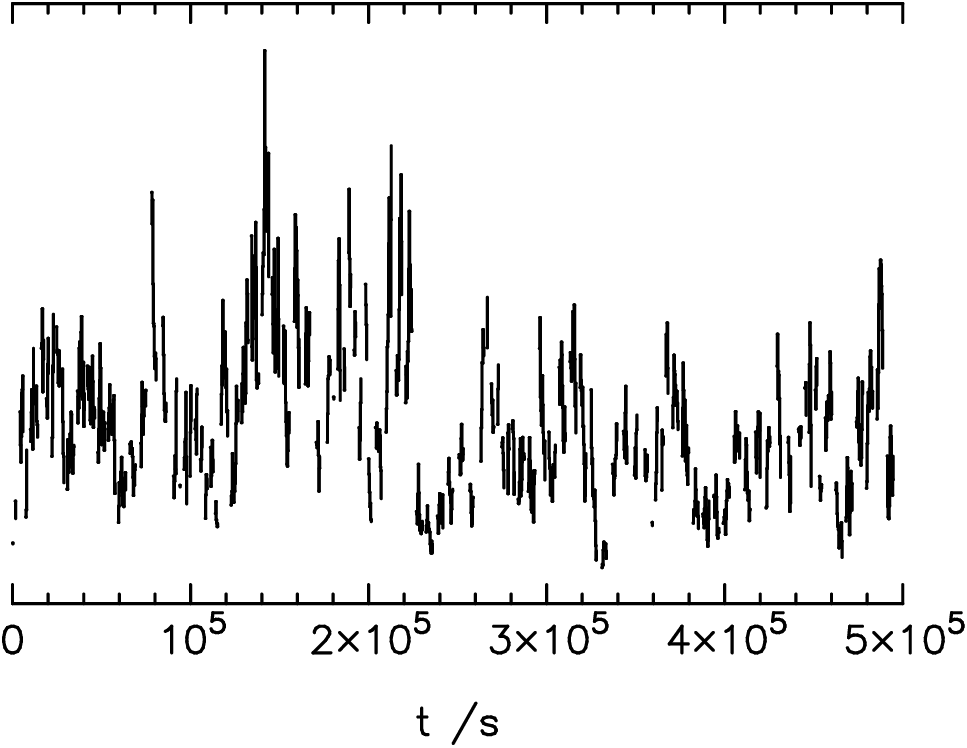}}
\hspace*{3mm}
\rotatebox{-90}{\includegraphics{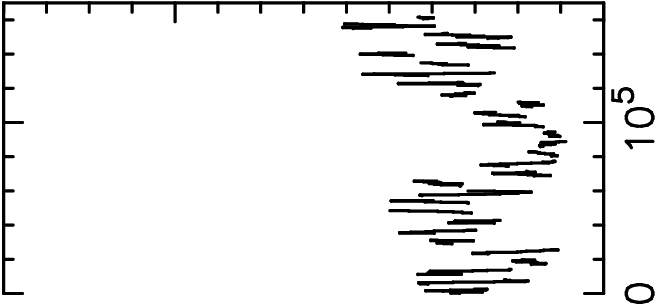}}
}
\caption{{\sc xis} light curves, 0.3-10\,keV,
for the three {\em Suzaku} observations, sampled at 256\,s: (left) 2005, (centre)
2008 Nov 6-12, (right) 2008 Nov 23-25.  The {\sc xis}\,0,2,3 observations of 2005 have
been scaled to the flux expected in two detectors, to allow comparison with the
{\sc xis}\,0,3 measurements in 2008.  Gaps indicate periods of no useful data.
Time values are set to zero at the first period of useful data in each observation.}
\label{fig:lightcurves}
\end{figure*}

The PCA shows the same generic features as the other type\,I AGN that it has been 
applied to (e.g. Mrk\,766, \citealt{miller07a}, MCG--6-30-15, \citealt{miller08a}).
The hard offset component dominates the {\sc pin} band, and 
contains atomic signatures, particularly of the Fe\,K edge and Fe\,K$\alpha$
emission line, indicative of the importance of atomic processes in shaping this
component.  Eigenvector-one is essentially a power-law, but with some
``warm absorber'' signatures imprinted on it in the soft band, and, unusually,
evidence of both an excess variable component above 20\,keV and weak evidence
for broadened Fe\,K$\alpha$ emission, with an equivalent width of about
$66_{-37}^{+43}$\,eV.  The width of the broad component is not well constrained,
but there is no
evidence for any excess variable component below 6\,keV, and in common with
other AGN, the ``red wing'' component is described by the constant offset
spectrum: the red wing feature in the 2008 data is implied by the PCA to be
simply a sum of the very hard-spectrum offset component and the power-law-like
eigenvector one.

The interpretation of eigenvector-one is fairly straightforward, as it implies
that the primary continuum has a power-law form of invariant index, whose amplitude
varies.  The amplitude variations could either be intrinsic to the source
producing the power-law, or could arise from changes in covering fraction
of an external absorber.  The variable hard excess could either be flux transmitted
through a partially-covering, high column density absorber, such as found in
PDS\,456 \citep{reeves09a} and 1H\,0419--577 \citep{turner09b}, or it could arise
as reflection from optically-thick, or nearly so, material. If there
are high column density partial-covering absorbers, we expect to see both transmitted
flux at high energy and Compton-scattered flux \citep[see ][]{miller09a}.

The offset component, which essentially describes
the source in its lowest observed flux state, may be explained either
as being dominated by reflection from optically-thick material, or as being the
residual signal that arises from partial-covering variations.  These possibilities
are discussed more by \citet{miller07a} and \citet{turner07a}, where the point
is made that it is fundamentally not possible to distinguish between these
possibilities by continuum fitting alone.  Although spectral models have
previously tended to be characterised as being dominated by either one or other
of absorption or reflection, it now seems likely that both physical processes
make significant contributions: absorbing zones of high column density also 
contribute a significant flux of Compton scattered light \citep{miller09a}.  
We return to discussion of the interpretation of the spectral variability
in section\,\ref{sec:discussion}, after consideration of the
variability on shorter timescales.

Finally, the Fe\,K$\alpha$ emission is particularly interesting\footnote{
\citet{turner09c} also discuss the companion line visible at 5.44\,keV in the offset
spectrum.}.
The narrow 6.4\,keV
line, produced by Fe\,{\sc i}-{\sc xvii}, 
appears only on the offset component, and hence
must have had an invariant flux in the 2005 and 2008 epochs. The implication is
that the gas responsible for the line experienced the same ionising continuum in
2005 as in 2008, despite the observed 7-10\,keV flux being a factor 2.5 higher
in 2008 than in 2005. We discuss later the interpretation that the dip into the 2005 low state was caused
by obscuration along our line of sight.

\subsection{Summary of key spectral variability results}
We summarise the key points from the above, to aid the later discussion.
\begin{enumerate}
\item NGC\,4051 shows the same general form of spectral variability as other type\,I
AGN, with a soft varying power-law and a less variable hard spectrum component
that creates a red wing below 6.4\,keV and {\sc pin}-band excess above the power-law
in the mean spectrum.
The 2005 low state is dominated by this hard component. The new analysis confirms 
the conclusions of \citet{pounds04c} and \citet{terashima09a}.
\item The 6.4\,keV\,Fe\,K$\alpha$ narrow emission line was
unchanged between 2005 and 2008 despite a significant change in continuum flux,
as also discussed by \citet{pounds04c}.
\item There is also a component of variable hard excess at $E \ga 20$\,keV that
varies with the power-law continuum on 20\,ks timescales, with evidence for
an associated Fe\,K$\alpha$ line component that is moderately broadened
(i.e. not relativistically broadened), by either Doppler or Compton scattering effects.
\end{enumerate}

\section{Power-spectrum and time delay measurement}
\subsection{Maximum-likelihood estimation of the power spectrum and time delays}
\label{sec:likel}
We now wish to complement the spectral variability analysis with analysis of the 
spectral variations in broader energy bands on shorter timescales.  In the case of
{\em Suzaku} observations it is not possible to use periodogram-type methods
\citep[e.g.][]{papadakis93a} because of the substantial gaps in the data caused by
the filtering of events and the short orbital timescale.  Table\,\ref{table:data} shows
that the fraction of usable data is in the range 0.48-0.55.  Furthermore, the
orbital period imposes regular gaps of period 5753\,s on the data.
Instead we use a maximum likelihood method, based on that of
\citet*{bond98a}, which is described more fully in the context of X-ray time series analysis
by Miller (in preparation).  The reader is referred to that paper for a full description
of the method, but briefly, we start by defining the power spectral density
(PSD), $P(\nu)$, as the Fourier transform of
the autocorrelation function, $\mathcal{A}(\tau)$,
\begin{equation}
\mathcal{A}(\tau) = \int_{-\nu_{\mathrm{max}}}^{\nu_{\mathrm{max}}}  \mathrm{d}\nu P(\nu) \cos(2\pi\nu\tau)\,
\mathrm{sinc}^2\left(\frac{\pi\nu}{2\nu_N}\right),
\label{eqn:1}
\end{equation}
where $\nu_N$ is the sampling Nyquist frequency (we assume uniform sampling in time) and
the $\mathrm{sinc}^2$ term arises from the effect of binning the counts in bins of
duration equal to the time sampling (see also \citealt{efstathiou01a}).
We do not construct a periodogram, but instead
use an iterative maximum likelihood method to find the best-fitting coefficients
for a model PSD given the measured time series.  We adopt a gaussian likelihood function,
$\mathcal{L}$, which thus takes the form
\begin{equation}
\mathcal{L} = \frac{1}{(2\pi)^{N/2}\left | C \right |^{1/2}}
\exp \left [-\frac{1}{2}\Delta^T C^{-1} \Delta \right ],
\label{eqn:likel}
\end{equation}
for a time series containing $N$ data samples,
where $C$ is a $N \times N$ model covariance matrix and $\Delta$ is the time series of fractional
fluctuations of length $N$. To find the best-fitting PSD we compute the expected covariance matrix
for an initial set of model parameter values using equation\,\ref{eqn:1} and iterate.
Because we can calculate the derivatives of the likelihood function with respect to
the model, from equation\,\ref{eqn:1}, we can use a Newton-type method as described by \citet{bond98a}, and 
the iterations typically converge within 5 iterations to a stable solution irrespective
of the number of model parameters.

The PSD model we adopt follows \citet{bond98a} in 
defining a set of ``bandpowers'', which we choose to be stepwise in $\nu P_\nu$:
\begin{equation}
\mathcal{A}(\tau) =
 2 \sum_i \int_{\nu_{1, i}}^{\nu_{2, i}} \frac{\mathrm{d}\nu}{\nu} (\nu_i P_i) \cos(2\pi\nu\tau)\,
\mathrm{sinc}^2\left(\frac{\pi\nu}{2\nu_N}\right)
\label{eqn:bandpowers}
\end{equation}
where each bandpower has PSD amplitude $P_i$ in the frequency range $\nu_{1, i}$--$\nu_{2, i}$.
Within each band the PSD is constant in $\nu P(\nu)$, which helps reduce
sharp steps in the PSD at bandpower edges compared with defining bandpowers uniform in $P(\nu)$.
As part of the likelihood maximisation procedure, we calculate the Fisher matrix for the PSD
coefficients, allowing us to have estimates of the errors and their covariance.  Those
errors include the sampling uncertainty that arises because we only observe a finite time series.
The likelihood estimation
is in the time domain, so the errors on the PSD coefficients are well-understood even in the
presence of a complex sampling window function.
The method is immune to gaps in the data, periodic or otherwise (if the time sampling
is such that one particular Fourier mode is badly sampled, this is simply reflected
in the corresponding uncertainty).  Multiple datasets spanning arbitrarily long time
baselines may be trivially combined, so we are able to analyse the three {\em Suzaku}
observations as either individual or combined time series.

The likelihood function that is maximised includes terms in the autocorrelation function
for both modes intrinsic to the source, described by $P(\nu)$, and Poisson shot noise.
Thus the likelihood method automatically corrects for Poisson noise in the data and is accurate
even in the low count rate regime.  
Because the shot noise is put into the likelihood model, 
the uncertainties on the PSD that we measure include
the statistical uncertainty arising from shot noise.
Although correction for high-frequency aliasing may also be included in this method (Miller, in preparation),
as both the high-frequency
part of the PSD has a steep spectrum, $P_\nu \propto \nu^{-\alpha}$ where $\alpha \simeq 2.1$
\citep{mchardy04a} and also aliased high-frequency modes are suppressed by the $\mathrm{sinc}^2$ term, 
the effects of aliasing are minimal and we make no aliasing
correction.

The likelihood method may be extended to simultaneously measure the PSDs of two time series, 
and their cross-spectrum frequency-dependent coherence and time delays (Miller, in preparation), 
and we show results from this analysis also. By incorporating time delays as additional
parameters in the likelihood model we can straightforwardly compute best-fit values and
errors.

\subsection{Red noise leak and the PSD bandpower widths}
\label{sec:redleak}
A well-known problem with any power-spectrum estimation method applied to finite time series
is that modes on timescales longer than the window function are not correctly measured.
In effect, long-timescale variations are interpreted instead as modes with periods on the scale
of the window function, a phenomenon known as ``red noise leak''.  If the variations don't have much
power on long timescales this may not be important, but as NGC\,4051 has a low frequency
PSD $P(\nu) \propto \nu^{-\alpha}$ where $\alpha \simeq 1$ at least down to $\nu \la 10^{-8}$\,Hz
\citep{mchardy04a}, red noise leak cannot be ignored. Miller (in preparation) has tested the effect of
red noise leak using data simulated with the \citet{mchardy04a} 
PSD of NGC\,4051 and sampled with the window function
provided by the 2008 {\em Suzaku} observations.  As expected, it is found that the lowest frequency
periods are measurably affected by red noise leak.  In principle in our PSD estimator we can include
low frequency modes by extrapolating the lowest bandpower down in frequency: however even this
is not a good solution as in the data we are not averaging over a large number of low frequency
modes, and furthermore the analysis becomes dependent on the model assumed for the extrapolation.

A related issue is that modes close in frequency are statistically correlated, because of the
finite window function.  We must therefore choose frequency widths for our bandpower estimates
that are broader than the window function to avoid unpleasant correlations and degeneracies
between neighboring bandpowers.  At the lowest frequency bandpowers, the effect of red noise leak
also introduces correlations and anti-correlations between neighbouring modes.  The best solution
is to choose bandpowers that are fairly broad.  In the following analysis we choose bandpowers
that are logarithmically spaced, with $\Delta\log_{10}\nu =0.4$ or $0.5$: at high frequencies
this is adequate to ensure that they are independent.  At low frequencies we bin together 
those bandpowers to make a single bandpower whose width is at least as large as the half-power
full-width (HPFW) of the window function.  We measure the HPFW by direct Fourier transform of
the window function, and finding the the frequency lower than which half the summed power is
found.  The lowest-frequency bandpower in our analysis thus tends to be extremely broad in
$\log\nu$.

Because of red noise leak, the PSD in this lowest bandpower should always be treated with caution in the 
subsequent analysis.  One particular concern which arises is in the measurement of time lags
between energy bands.  A long period mode which has some small time shift between two energy
bands will, in the low frequency limit, 
simply appear as having a shift in the mean value between the bands within the observed
time window.  The shift in the mean will be removed because we use the 
observed mean value to convert the measured time series into fractional fluctuations $\Delta$,
and hence the time lag will disappear.  We return to this concern in sections\,\ref{sec:lags}
and \ref{sec:reverb}.

\begin{figure}
\resizebox{0.45\textwidth}{!}{
\rotatebox{-90}{
\includegraphics{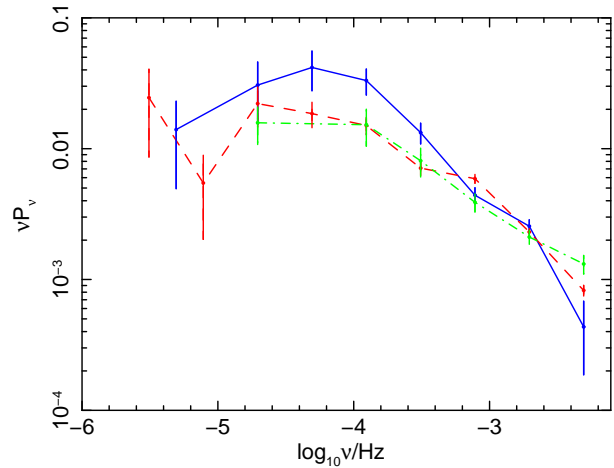}
}}
\caption{Comparison of the PSDs for NGC\,4051 derived from the three epochs: 2005 
(solid curve),
2008\,Nov\,6-12 (dashed), 2008\,Nov\,23-25 (dot-dashed), 
showing $\log_{10}\nu P_{\nu}$ against
$\log_{10}\nu/{\rm Hz}$.  Points with errors bars are plotted at the midpoint in $\log\nu$
of each bandpower, error bars show the Fisher matrix uncertainties, lines are drawn to
connect the points in each epoch of observation.}
\label{fig:obs3comp}
\end{figure}

\begin{figure*}
\resizebox{0.45\textwidth}{!}{
\rotatebox{-90}{
\includegraphics{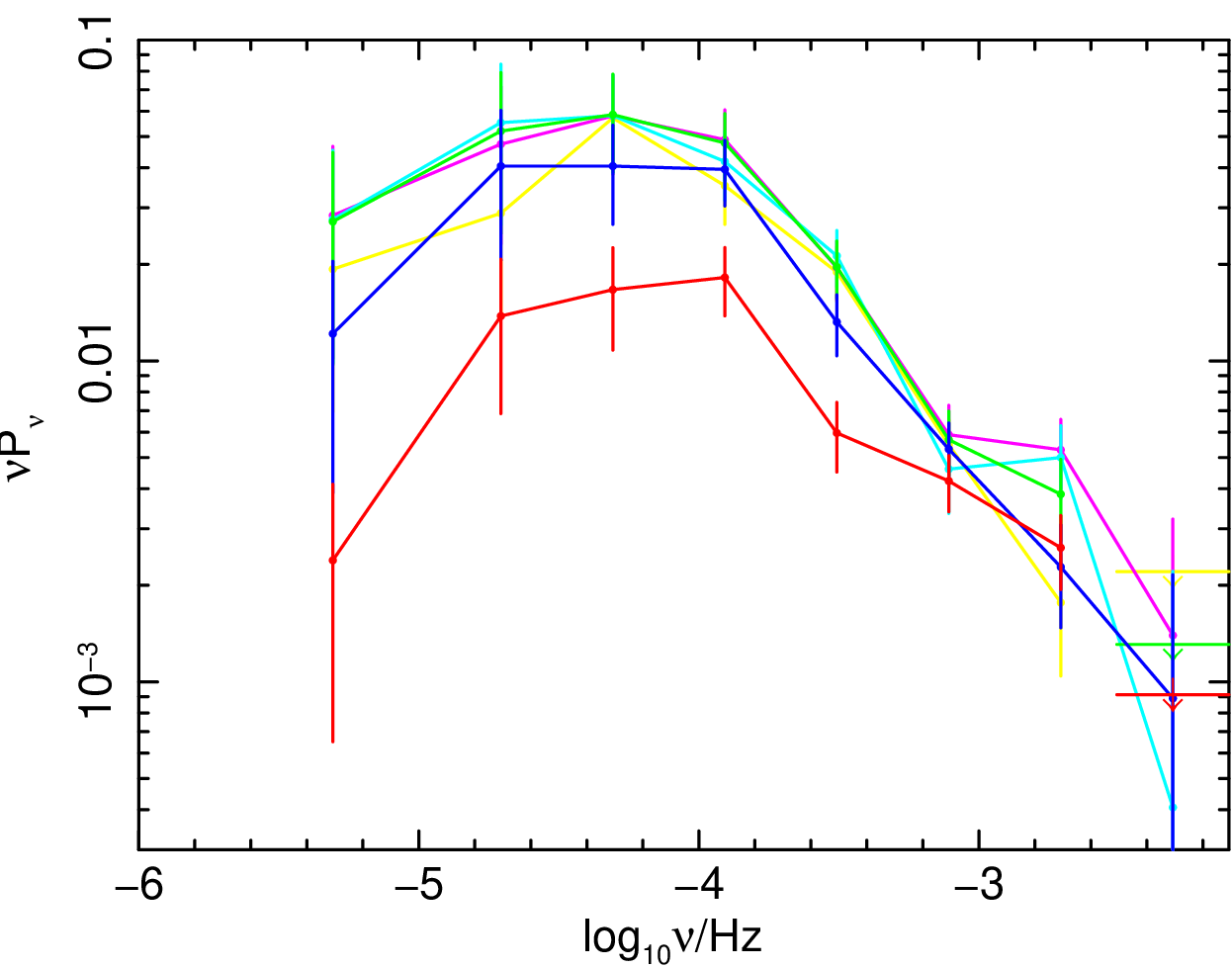}
}}
\hspace*{5mm}
\resizebox{0.45\textwidth}{!}{
\rotatebox{-90}{
\includegraphics{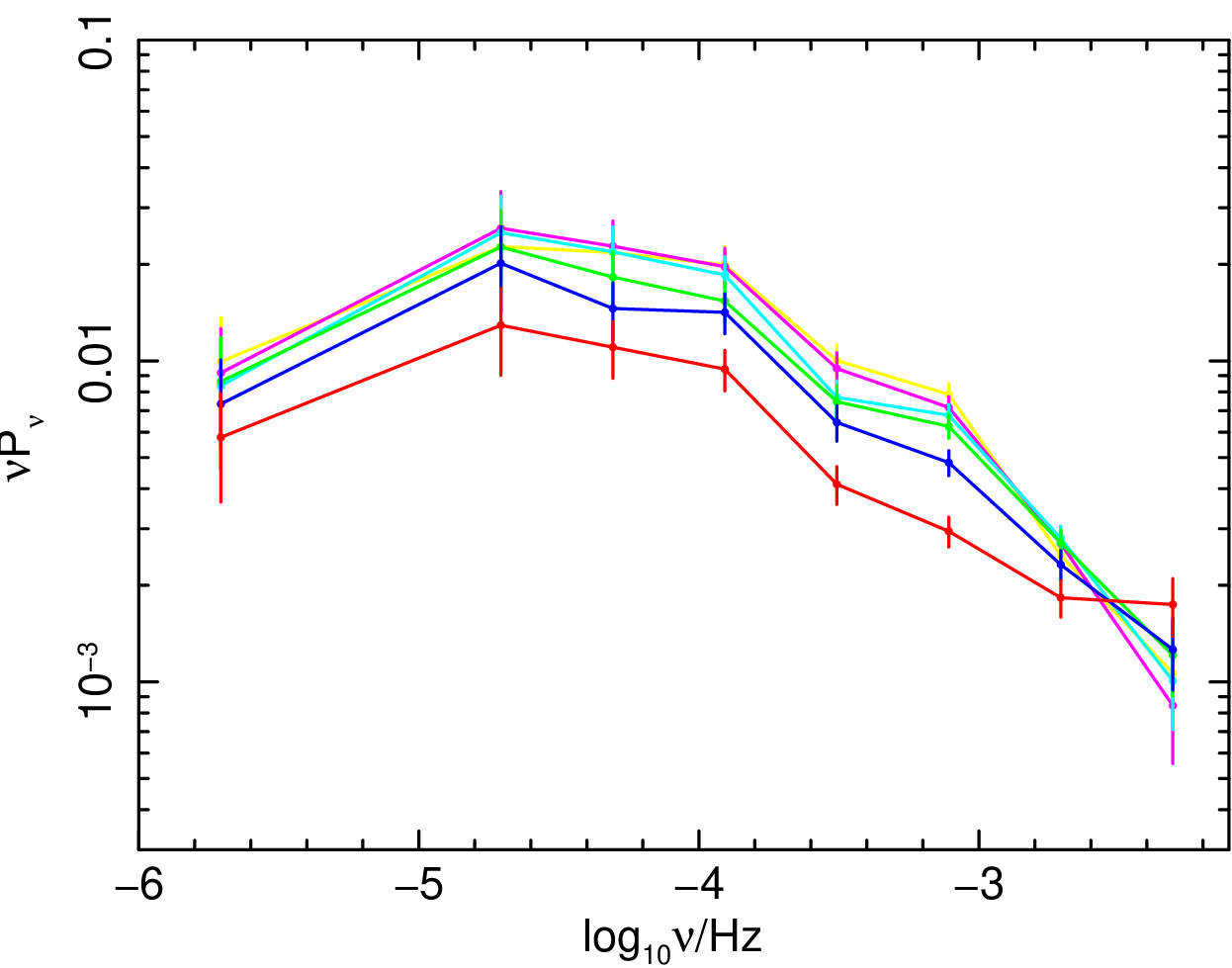}
}}
\caption{PSDs, plotted as $\nu P(\nu)$ 
from the observations of NGC\,4051 in 2005 (left) and 2008 (right).
There are six energy bands:
0.3-0.94 (yellow),
0.94-1.2 (magenta),
1.2-1.5 (cyan),
1.5-2.1 (green),
2.1-3.8 (blue),
3.8-10\,keV (red).  
The Fisher matrix uncertainties are shown as error bars.
The PSD in the highest frequency bandpower in 2005 was consistent with zero in
three of the energy bands, the plot shows the corresponding 2$\sigma$ upper limits.
}
\label{fig:psd}
\end{figure*}

\subsection{Results}

\subsubsection{The time series}
Time series were constructed from the data by binning and sampling at regular intervals.
Any time bins when the source observation covered less than half a bin were
not used: the remainder were corrected by the ``on-time'' in each bin.  
As an illustration of the variability, Fig.\,\ref{fig:lightcurves} shows
the full band, 0.3-10\,keV, time series sampled at 256\,s 
(for the PSD determination in each band, sampling times of 64\,s were used).  
The frequent and periodic gaps in the data are clearly visible.  The variability is
clear however, and even by eye it can be seen that it appears to have different
characteristics in 2005 from the two observations in 2008.  We quantify this in the
next section.

\subsubsection{Non-stationarity of the variations}
The first {\em Suzaku} observation was obtained three years before the second two observations,
when the source was in a substantially lower flux state.  Even the second and third observations,
taken two weeks apart, have different mean flux (section\,\ref{sec:data}).
We therefore start by comparing
power spectra from the three observations to test whether they are consistent with stationary
behaviour, or whether there are differences.  Fig.\,\ref{fig:obs3comp} shows PSDs measured
from each {\sc xis} dataset using the energy range 0.3-10\,keV, although because of the steep
source spectrum and the higher effective area in the soft band, the measured spectrum is dominated
by the soft band.  The time sampling used was 64\,s, with a bandpower interval of 
$\Delta\log_{10}\nu = 0.4$.  Bandpower intervals smaller than the window function full width (as
defined in section\,\ref{sec:likel}) were coadded as described above.
The minimum frequencies differ for each dataset owing to
the differing lengths of observation (Table\,\ref{table:data}).

The PSDs at low frequencies, $\nu \la 10^{-5}$\,Hz, show significant variation owing to 
uncorrected red noise leak.  Above this frequency however useful comparison may be made.
The two observations from 2008 are closely consistent with each other, the third observation
having only a marginally significant excess of power at the highest frequency bandpower.
The 2005 data appear significantly different however, with more power in the mid frequency
range and a steeper spectrum to high frequencies.

The mean levels used to normalise the fluctuations were different in the three datasets of course.
If we were to scale observations 2 and 3 to the same mean, their amplitudes would differ.
This implies either what has previously been suggested \citep{uttley01a, uttley05a}, 
that the fluctuations are intrinsically ``fractional'' in nature, or that some intrinsic
fluctuations have been modulated by a secondary effect such as a varying multiplicative
absorption.  
The 2005 observation, however, clearly has different shape.
To quantify this, we measure the slope of the PSD at high frequencies by fitting 
a linear relation between $\log P_\nu$ and $\log\nu$ for
$\nu > 10^{-4}$\,Hz.  Such a functional fit is a good fit for the 2005 and 2008 Nov\,23-25 PSDs
but fits less well the 2008\,Nov\,6-12 PSD: as the aim of this exercise is to see whether there
are statistically-significant differences between the PSDs of the three epochs, this simple
parameterisation is sufficient.  A minimum-$\chi^2$-squared fitting method is adopted, where as
well as taking account of the statistical uncertainty on each of the fitted PSD points, we
also include the covariance between the points as determined from the Fisher information
matrix.  We find best-fit slopes $\alpha = 2.00 \pm .08$, $1.78 \pm .03$ and $1.67 \pm .08$,
for each of 2005, 2008\,Nov\,6-12 and 2008\,Nov\,23-25 respectively,
where $P_\nu \propto \nu^{-\alpha}$.  Thus the two observations in 2008 have high-frequency slopes
that are consistent with each other, but the 2005 PSD has a steeper high-frequency slope,
that differs by $2.7\sigma$ from the 2008\,Nov\,6-12 slope and by $3\sigma$ from the 2008\,Nov\,23-25
slope.

We return in section\,\ref{sec:discussion} 
to discussion of these points in the full context of all the results, but for now we
note that the PSD behaviour of the source appears non-stationary, given the different
shape PSDs between 2005 and 2008, and in the subsequent PSD analysis we treat the 2005 and 2008
data as separate time series.

\subsubsection{Energy-dependent PSDs}
\label{sec:psd}
To search for any dependence of PSD on photon energy, we divided the {\sc xis} photon events into
six bandpasses covering the range 0.3-10\,keV.  
The likelihood method corrects for shot noise in the data, but as we shall
be looking for differences between the PSDs of the six bandpasses, we have chosen energy
boundaries such that each band has the same count rate.  
The source photon spectrum is steep, and
the {\sc xis} effective area peaks in the soft band, so inevitably this 
means the bands are not at all spaced evenly in energy.  The energy bands were 
0.3-0.94, 
0.94-1.2, 
1.2-1.5, 
1.5-2.1, 
2.1-3.8, 
3.8-10\,keV.

The results are shown in Fig.\,\ref{fig:psd} for a time sampling of 64\,s.
As found by \citet{mchardy04a}, and as expected from
the longer timescale spectral variability in section\,\ref{sec:pca}, the harder bands have less
overall variability.  However, it is also clear that the hardest energy band, 3.8-10\,keV, has
a significantly flatter PSD than the softer bands, in fact having the highest power of any energy band
at the highest frequency measured, despite having significantly lower power at lower frequencies,
although given the error bars we cannot say that the hard band has significantly more power than
the softer bands.  This result is in good agreement with the model-fitting approach of
\citet{mchardy04a} applied to the 2001 {\em XMM-Newton} data, where evidence for flattening of the
PSD with energy was also found.  Overall, the shape and normalisation of the {\em Suzaku} PSD is
in excellent agreement with that {\em XMM-Newton} PSD.  There is some evidence that the lowest
frequencies dip below the best-fit relation $P(\nu) \propto \nu^{-1}$ found by \citet{mchardy04a},
but the departure is consistent with the uncertainty in the \citeauthor{mchardy04a} slope
and in any case may be affected in our analysis by red noise leak.  In the following
analysis we ignore the PSD shape at $\nu < 10^{-5}$\,Hz.

\subsubsection{Time delays and coherence}
\label{sec:lags}
We can also search for evidence of time lags between energy bands, a phenomenon which has
now been detected in a number of AGN, as discussed in the introduction.  Again, we follow
the method described by Miller (in preparation) where the cross power spectrum between two bands 
is calculated in terms of some amplitude and a time lag between the bands.  The value
of time lag derived corresponds exactly to the time lag derived from the cross-correlation
function in the case where every frequency bin is constrained to have the same lag.  By
working in the Fourier domain however we can also investigate frequency-dependent lags,
effectively being the lags obtained by filtering the cross-correlation function with a top-hat
filter in frequency.  The lags we derive by this method are directly comparable to those
derived by other authors \citep[e.g.][]{mchardy04a}, but they have been arrived at by the
maximum likelihood method rather than periodogram or Monte Carlo methods.

For this exercise we combine all datasets to maximise the signal-to-noise in the time lags.
Even though the shape of the PSD varies between 2005 and 2008, the derived time lags
show no significant evidence for epoch dependence: the 2008 data alone yield time lags very
close to those from the combined dataset, with slightly larger uncertainties.  The 2005 data
alone yield time lags consistent with the 2008 data but with large uncertainties.
As found by \citet{mchardy04a}, we find significant time lags between energy bands, where 
the harder band lags the softer band and where the lag increases with energy difference.
We define the energy bands similarly to section\,\ref{sec:psd}.
Again, the bands have to be broad in energy in order to obtain an adequate count rate,
but to best investigate the energy dependence of the time lags 
we relax the requirement to have equal count rates in all energy bands, and the 
hardest band from section\,\ref{sec:psd} is split into two, 3.8-6 and 6-10\,keV.
The time sampling chosen was 256\,s: this longer time sampling improves the count statistics
in each time bin, and we shall see from the results that we are most interested in time lags
on periods longer than 1\,ks and with values for the lag measured in hundreds of seconds.
Accurate characterisation of the measurement errors is important for the time delays, so here we
supplement the standard Fisher matrix error measurements with direct measurement of the likelihood
distribution, obtained by stepping through a range of time delays at each frequency, while maximising the
likelihood by allowing all other coefficients to vary.  We define two
confidence intervals by the regions $2\Delta\ln\mathcal{L}\le 1$ and $2.7$ respectively: as the
likelihood distributions appear to be close to gaussian these correspond closely to the standard
68-percent and 90-percent confidence regions of a normal distribution.  The 68-percent confidence
regions are also close to the Fisher matrix errors, again as expected for gaussian uncertainties.
Table\,\ref{table:lags} gives the Fisher matrix and 68- and 90-percent confidence regions for each
frequency band.  

\begin{figure}
\resizebox{0.45\textwidth}{!}{
\rotatebox{-90}{
\includegraphics{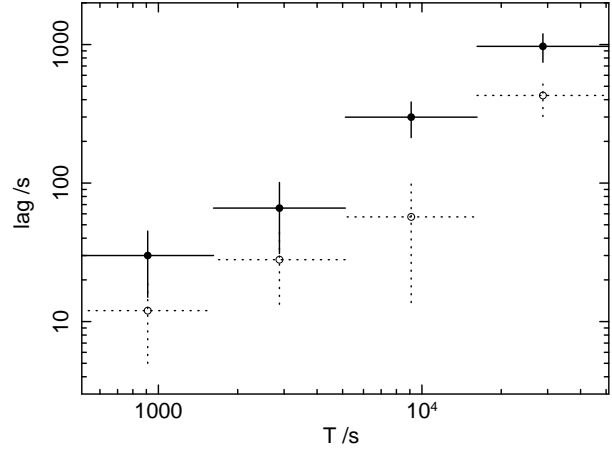}
}}
\caption{Time lags between the 0.3-0.94\,keV band and the 
6-10\,keV (solid points and error bars) and 2.1-3.8\,keV (open points and dotted error
bars) bands,
as a function of time
period $T$ for the combined 2005-8 dataset.
Vertical error bars show the 68\,percent confidence regions. 
Horizontal bars show the range of time periods included in
each bandpower.}
\label{fig:lags}
\end{figure}

In these results, lags up to 970\,s are detected with respect to the softest band, 
that increase with band energy and with the period of the fluctuations, as found by
\citet{mchardy04a}.
The longest period point is consistent with zero lag but with a very large uncertainty.  Even so,
it does appear discrepant compared
with the results of \citet{mchardy04a} and is worth further consideration.
\citeauthor{mchardy04a} found a lag of 3000\,s on timescales $\sim 70$\,ks between the 0.1-0.5\,keV
and 5-10\,keV bands.  The disagreement with the 90\,percent confidence upper limit of 600\,s
(Table\,\ref{table:lags}) could indicate a variation in source structure between 2001 and 2008:
a more likely possibility, however, is that the long timescales have not been sufficiently sampled 
by the {\em Suzaku} observations to allow accurate measurement, in contrast with the {\em RXTE}
observations analysed by \citet{mchardy04a} that spanned 6 years.  As discussed in section\,\ref{sec:redleak},
red noise leak into the longest period bin may corrupt the measurement of the time delay.
We therefore exclude this bin from further consideration.
Time lags between the 0.3-0.94\,keV band and the 
2.1-3.8 and 6-10\,keV bands are shown in Fig.\,\ref{fig:lags}, plotted logarithmically (other
bands are not shown, for clarity, but may be inspected in Table\,\ref{table:lags}).

\begin{figure}
\resizebox{0.45\textwidth}{!}{
\rotatebox{-90}{
\includegraphics{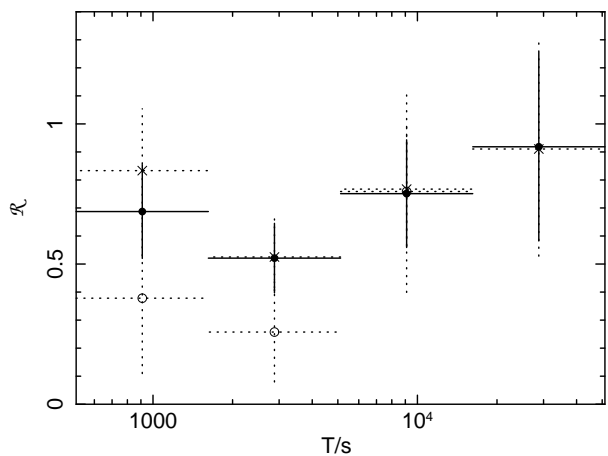}
}}
\caption{Coherence $\mathcal{R}$ 
between the 0.3-0.94\,keV and 6-10\,keV bands as a function of time
period $T$ for the combined 2005-2008 dataset (solid symbols).  
Vertical error bars show the Fisher matrix errors, horizontal bars
show the range of time periods included.  
Also shown are the coherence between the same bands for the 2005 data
alone (open symbols and dashed error bars) and for the 2008
data alone (cross symbols and dashed error bars).
The highest time period in the 2005
dataset is not shown as the error bounds are too large to be useful.
}
\label{fig:coherence}
\end{figure}

\begin{table}
\caption{Time lags between the 0.3-0.94\,keV band and bands at higher energy, as a function of time
period $T$ for the combined 2005-8 dataset.
Columns give the range of periods in each band and the lag, followed by
three estimates of the errors: Fisher matrix, direct evaluation 68\% confidence region
and direct evaluation 90\% confidence region.  The longest time period in each energy band
is likely affected by red noise leak and is not used in the analysis.
}
\begin{tabular}{llrrrr}
$T$ & range    & lag & Fisher & 68\% & 90\% \\
/ks & /ks & /s & error /s & error /s & error /s \\ 
\hline \\
\multicolumn{6}{l}{{\bf 6-10\,keV}}\\
2879    & 51.2 - 162000 & $-564$   & $\pm 913$ & $^{+694}_{-715}$ & $^{+1163}_{-1190}$ \\
28.8    & 16.2 - 51.2   & 970    & $\pm 212$ & $\pm 225$ & $^{+372}_{-374}$ \\
9.1     & 5.12  - 16.2  & 299    & $\pm 84$ & $\pm 86$   & $\pm 142$ \\
2.9     & 1.62  - 5.12  & 66     & $\pm 35$ & $\pm 35$   & $\pm 58$  \\
0.91    & 0.512 - 1.62   & 30     & $\pm 15$ & $\pm 17$   & $\pm 28$  \\
\hline \\
\multicolumn{6}{l}{{\bf 3.8-6\,keV}}\\
2879    & 51.2 - 162000 & $-872$   & $\pm 711$ & $^{+558}_{-563}$ & $^{+932}_{-942}$ \\
28.8    & 16.2 - 51.2   & 700    & $\pm 180$ & $\pm 193$ & $\pm 320$ \\
9.1     & 5.12  - 16.2  & 94    & $\pm 59$ & $\pm 58$   & $\pm 95$ \\
2.9     & 1.62  - 5.12  & 57     & $\pm 24$ & $\pm 24$   & $\pm 39$  \\
0.91    & 0.512 - 1.62   & 30     & $\pm 9$ & $\pm 10$   & $\pm 16$  \\
\hline \\
\multicolumn{6}{l}{{\bf 2.1-3.8\,keV}}\\
2879    & 51.2 - 162000 & $-880$   & $\pm 566$ & $^{+474}_{-416}$ & $^{+767}_{-713}$ \\
28.8    & 16.2 - 51.2   & 429    & $\pm 124$ & $\pm 133$ & $\pm 219$ \\
9.1     & 5.12  - 16.2  & 57    & $\pm 44$ & $\pm 44$   & $\pm 72$ \\
2.9     & 1.62  - 5.12  & 28     & $\pm 16$ & $\pm 16$   & $\pm 26$  \\
0.91    & 0.512 - 1.62   & 12     & $\pm 7$ & $\pm 7$   & $\pm 12$  \\
\hline \\
\multicolumn{6}{l}{{\bf 1.5-2.1\,keV}}\\
2879    & 51.2 - 162000 & $-420$   & $\pm 477$ & $^{+398}_{-351}$ & $^{+645}_{-600}$ \\
28.8    & 16.2 - 51.2   & 373    & $\pm 101$ & $\pm 106$ & $\pm 176$ \\
9.1     & 5.12  - 16.2  & 57    & $\pm 35$ & $\pm 35$   & $\pm 57$ \\
2.9     & 1.62  - 5.12  & 7     & $\pm 13$ & $\pm 13$   & $\pm 22$  \\
0.91    & 0.512 - 1.62   & 2     & $\pm 6$ & $\pm 6$   & $\pm 11$  \\
\hline \\
\multicolumn{6}{l}{{\bf 1.2-1.5\,keV}}\\
2879    & 51.2 - 162000 & $-444$   & $\pm 441$ & $^{+343}_{-366}$ & $^{+582}_{-601}$ \\
28.8    & 16.2 - 51.2   & 287    & $\pm 101$ & $\pm 106$ & $\pm 176$ \\
9.1     & 5.12  - 16.2  & 48    & $\pm 33$ & $\pm 32$   & $\pm 52$ \\
2.9     & 1.62  - 5.12  & $-1$     & $\pm 12$ & $\pm 12$   & $\pm 20$  \\
0.91    & 0.512 - 1.62   & 12     & $\pm 5$ & $\pm 5$   & $\pm 9$  \\
\hline \\
\multicolumn{6}{l}{{\bf 0.94-1.2\,keV}}\\
2879    & 51.2 - 162000 & $-427$   & $\pm 406$ & $^{+317}_{-339}$ & $^{+537}_{-558}$ \\
28.8    & 16.2 - 51.2   & 216    & $\pm 108$ & $\pm 114$ & $\pm 189$ \\
9.1     & 5.12  - 16.2  & 49    & $\pm 28$ & $\pm 27$   & $\pm 45$ \\
2.9     & 1.62  - 5.12  & 10     & $\pm 11$ & $\pm 11$   & $\pm 19$  \\
0.91    & 0.512 - 1.62   & 6     & $\pm 5$ & $\pm 5$   & $\pm 9$  \\
\hline
\end{tabular}
\label{table:lags}
\end{table}

We also investigated time lags between the {\sc pin} and {\sc xis} bands, however the {\sc pin}
data were too noisy for any accurate time lag to be measured (although 
were consistent, within the large uncertainties, with the lags for the hardest band
shown in Table\,\ref{table:lags}).

Finally, we investigate the coherence, $\mathcal{R}$, between these two energy bands, where coherence is defined
following \citet{nowak96a} as the ratio of the cross-spectrum amplitude squared to the product
of the individual PSD amplitudes for the two bands, as a function of frequency.
It is defined to lie in the range $0 \le \mathcal{R} \le 1$.  High coherence
implies the different wavebands are highly correlated, low coherence implies they are not.
Fig.\,\ref{fig:coherence} shows $\mathcal{R}$ again as a function of frequency, showing only the Fisher
matrix errors.  The errors are large, so we have not directly measured the likelihood distributions
in this case (because we only plot the Fisher errors, the error bars are not constrained to
lie within the range $0 \le \mathcal{R} \le 1$).  Results from the highest time period in 2005 are
not shown as the uncertainty in $\mathcal{R}$ is of order unity.  In the combined dataset,
2005-8, and in the 2008 data alone,
the coherence is high at all frequencies.  However, investigating the 2005 low state
data alone, the coherence drops for short period variations.

\subsection{Summary of key timing analysis results}
We also now summarise the key timing analysis results.
\begin{enumerate}
\item The PSDs demonstrate non-stationary behaviour between 2005 and 2008.  
\item The PSD measured for the hard band is significantly lower in amplitude
and with a flatter frequency-dependence than the soft band PSDs, as previously found
by other authors.  The difference between bands is more marked in 2005 than in 2008.
\item We confirm the existence of both frequency-dependent lags of harder-energy bands 
with respect to softer-energy bands, up to $970 \pm 225$\,s, and the energy dependence
of the lags.
\item The coherence between soft and hard bands is high in the 2008 data, but is
low in the 2005 data for short period variations.
\end{enumerate}

\section{Discussion}
\label{sec:discussion}
\subsection{Comparison of timing results with previous observations}
The power spectrum and frequency-dependent time lags measured from the {\em Suzaku} observations
of NGC\,4051 are in remarkably close agreement with the measurements reported by
\citet{mchardy04a}, despite being based on data taken with different observatories
({\em Suzaku} v. {\em XMM-Newton} and {\em RXTE}) on widely separated dates (2005-8 v. 1996-2002)
and using different analysis methods (maximum likelihood versus Monte Carlo).  The key
results that are in common are:
\begin{enumerate}
\item NGC\,4051 is overall less variable at high energy than at low energy over most
of the frequency range.
\item The 
high-energy PSD is flatter so that by time periods about 128\,s the variability in each band
is similar.
\item The softer bands in particular show a sharp break in the power spectrum at 
$\sim 10^{-4}$\,Hz.
\item Frequency-dependent time lags are found, with hard band lagging the soft band, with
lag increasing with energy difference, and lag increasing with period.  The maximum lag
found in the {\em Suzaku} analysis is $970 \pm 225$\,s for periods of 10s of ks, 
similar to the maximum lag found by \citet{mchardy04a} on similar periods
of approximately 800\,s.  
\end{enumerate}
There are some differences:  \citet{mchardy04a} find the lag continues to increase with period,
which may be attributable to the much longer time baseline afforded by the 6 years of {\em RXTE}
data they included; and they found a more significant decrease in coherence with decreasing 
period which may be due partly to the difference in analysis methods and partly to the inclusion
of low-state data in the \citeauthor{mchardy04a} analysis.
By analysing the 2005 low
state separately from the 2008 higher state, we find significant differences in the PSDs
and significantly lower coherence between bands in 2005.

\subsection{Reverberation time delays}
\label{sec:reverb}
The frequency-dependent time delays in this and other AGN have previously been interpreted in terms
of a model of intrinsic variations hypothesised to explain those features,
comprising perturbations propagating inwards on an accretion disk where harder X-ray emission
is emitted from smaller radii \citep{arevalo06b}.  While such models are not ruled out by the analysis
presented here (the results largely agree with previous analyses), they do not have any predicted
relationship to the PCA spectral decomposition that we have seen in section\,\ref{sec:pca}, and
although the earliest results indicated lags that varied linearly with time period, more
recent results reveal a more complex dependence of lag on period that has been interpreted as
indicating the presence of multiple zones of fluctuating propagations \citep{mchardy07a}.
However, there is a natural
explanation of the frequency-dependent time delays in terms of reverberation delays
\citep[e.g.][]{peterson93a} which we can relate directly to the PCA spectral components.
In section\,\ref{sec:pca} we saw the existence of a hard component
spectrum that dominates the low state, in common with many other AGN.  If this component has a
significant contribution from reflection, this may cause the observed time delays.  In fact, 
reverberation models predict just the sort of frequency-dependent time delays that
are observed here.  We illustrate this with a simple model.

\begin{figure}
\resizebox{0.45\textwidth}{!}{
\rotatebox{-90}{
\includegraphics{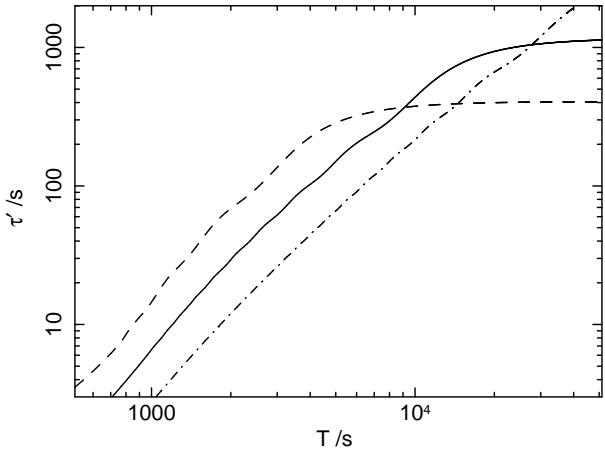}
}}
\caption{Predicted time lags, $\tau^\prime$, from the simple thick-shell 
reverberation model as a function of fluctuation
time period $T$ for $\tau_{\rm max}=10^{3.5}$\,s (dashed curve), $10^4$\,s (solid curve) and $10^{4.5}$\,s
(dot-dash curve).  The reflection factor $f=0.8$ for all curves.
}
\label{fig:lagmodels}
\end{figure}

Let us suppose that at any frequency, the continuum flux we observe comprises two components, one
a component from a central source that is seen directly, plus a time-delayed component reflected
from distant material.  Consider one Fourier mode of the measured source variability with angular
frequency $\omega$,
\begin{equation}
a_\omega^\prime e^{i\omega(t-\tau^\prime)} =
a_\omega e^{i\omega t} + a_\omega f \int_0^{\tau_{\rm max}} b_\tau e^{i\omega(t-\tau)} \mathrm{d}\tau,
\label{eqn:reflectedmode}
\end{equation}
where $a_\omega$ is the amplitude of that mode, $\tau$ is the time delay of reflected emission from some region,
primed quantities denote the observed amplitude and time delay in the combined direct plus reflected signal,
$b_\tau$ is the transfer function defined over
the range $0 \le b_\tau \le \tau_{\rm max}$ and $f$ is the total reflected fraction of light. We expect
$0 \le f < 1$ unless the illuminating source is obscured. 
It can be seen that the phase of the combined signal, contained in the term $e^{-i\omega\tau^\prime}$,
is related to the Fourier transform of the transfer function \citep[see related results discussed by][]{peterson93a}.
A simple case is given by considering
reflection from a thin spherical shell of radius $r$: the transfer function is then a uniform probability
distribution over the range $0 \le \tau \le \tau_{s}$ where $\tau_{s} = 2r/c$, and the
imaginary part of equation\,\ref{eqn:reflectedmode}, 
that is responsible for the observed time lags, is oscillatory.  
Smooth time lag functions may be produced by thick shells of reflection.  A thick 
spherical shell, with uniform reflected light per unit radius within a range $r_1 \le r \le r_2$,
has an observed phase term given by
\begin{equation}
\frac{a_\omega^\prime}{a_\omega} e^{-i\omega\tau^\prime} = 
1-i\frac{f}{\omega(\tau_{\rm max}-\tau_1))}\int_{\tau_1}^{\tau_{\rm max}}\frac{\mathrm{d}\tau_s}{\tau_s}(1-e^{-i\omega\tau_s})
\label{eqn:lagmodels}
\end{equation}
where we integrate over the thickness of the shell, defined by $\tau_1 = 2r_1/c$ and $\tau_{\rm max}=2r_2/c$.
Fig.\,\ref{fig:lagmodels} illustrates the predicted delays, $\tau^\prime$, as a function of the fluctuation time
period $T=2\pi/\omega$ for $f=0.8$, $\tau_1=500$\,s and a range of values of $\tau_{\rm max}$.
The frequency-dependence arises because high frequency modes are averaged out by the transfer
function, and the remaining variability is thus more dominated by the illuminating source alone.
Echoes of lower frequency modes can survive in the reflected signal, resulting in a measurable delay.
The suppression of high-frequency modes in the reflected spectrum has an effect on the PSD,
with the power in
modes at $\omega \ga 1/\tau_{\rm max}$ being suppressed by a factor $(1+f)^{-2}$ with respect
to those at lower frequencies.  The coherence between the direct signal and the combined direct plus
reflected signal tends to unity at low and high frequencies and varies as $\cos^2(\omega\tau^\prime)$
at intermediate frequencies.

One particular point to note is that the measured time lags are significantly less than the
light travel time to the reflecting region, because the coaddition of primary and reflected
light dilutes the time lag signature. In the limit $\omega \rightarrow 0$, the observed
time lag tends to a maximum, 
\begin{equation}
\tau^\prime \rightarrow \frac{f}{4(1+f)}\frac{(\tau_{\rm max}^2-\tau_1^2)}{(\tau_{\rm max}-\tau_1)},
\end{equation}
as seen in Fig.\,\ref{fig:lagmodels}
at high time periods for the models with low $\tau_{\rm max}$, so $\tau^\prime < \tau_{\rm max}$ always,
and in estimating the distance of the reverberating zone from the illuminating source we must allow
for this dilution of the lag time.  

It would be interesting if a signature of a maximum time lag could be found in the data.
In NGC\,4051 there is no evidence that such
a maximum has been found, either in this analysis or in that of \citet{mchardy04a}.  However,
a maximum lag has been found in Ark\,564 \citep{arevalo06a, mchardy07a}, lending support to
the reverberation explanation.  Another feature of the Ark\,564 analysis is that at the highest
frequencies there is some evidence for negative time lags 
(i.e. hard band leading soft band, \citealt{mchardy07a}) -
in the reverberation model this may arise at high frequencies as a result of the 
time-lag oscillatory behaviour produced by a reflecting region that only partially covers the source.
In effect, such apparent negative lags are an artefact created by the Fourier decomposition
of the time lag signature.

\begin{figure}
\resizebox{0.45\textwidth}{!}{
\rotatebox{-90}{
\includegraphics{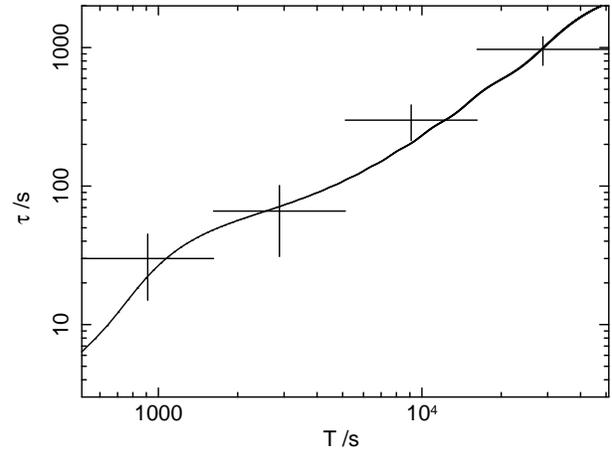}
}}
\caption{Predicted time lags, $\tau^\prime$, from the two-component thick-shell 
reverberation model, with parameter values as described in the text, 
as a function of fluctuation time period.  Also shown are the measured hard-band 
lag values from Fig.\,\ref{fig:lags}.  Note that this model is illustrative only and
has not been fitted to the measured values.
}
\label{fig:doubleshell}
\end{figure}

To measure $\tau^\prime$ in data, we need to be able to compare the combined, direct plus reflected,
signal with the direct signal alone.
In our measurements we find a lag between hard-band photons and soft-band photons.  
If we assume that in the soft-band we are seeing little reflected light, but that the hard
band signal is a combination of direct and reflected light,
we may compare
Fig.\,\ref{fig:lags} with Fig.\,\ref{fig:lagmodels}.
It can be seen that there is good qualitative agreement between the simple reverberation model
and the observed time lags, although the time lag for the shortest periods is larger than expected.  
The profile of the frequency-dependence (the ``lag spectrum'') depends on the assumed transfer
function: a fall-off in reflection with radius as $1/r$, in either spherical or flattened
geometries, tends to produce a flatter relation between lag and period.  
As a simple example of a more complex transfer function,
we consider a two-component reflection model, in which an inner reflecting zone has parameters
$f=0.3$, $\tau_1=200$\,s, $\tau_{\rm max}=800$\,s and a second zone extending to larger radii
has parameters $f=0.8$, $\tau_1=800$\,s, $\tau_{\rm max}=30$\,000\,s, where these parameter values
have been chosen purely as an illustration and have not been obtained from fitting to the
data (there are three parameters that primarily affect the shape of the lag spectrum, and
only four data points, so we do not consider fitting of this model to be particularly useful
for this dataset).  In this particular illustration, the total reflected light exceeds the
light that is directly received by the observer, implying some loss of direct light by either
absorption or scattering.
The resulting predicted lag spectrum is shown in Fig.\,\ref{fig:doubleshell}
together with the measured hard-band time lags from Fig.\,\ref{fig:lags}.
Such additional complexity in the reverberation model is able to reproduce more complex
lag spectra, leading to the hope that improved datasets may be able to better constrain
the distribution of circumnuclear material.
Although accurate
measurement of the lag spectrum could allow the transfer function to be determined, 
the present quality of data are insufficient to allow meaningful
determination of the transfer function shape, and despite the lag excess on short 
timescales, overall the simple
thick-shell model provides a reasonable description of the present dataset, and we now 
investigate the consequences of that simplest model.

\begin{figure}
\resizebox{0.45\textwidth}{!}{
\rotatebox{-90}{
\includegraphics{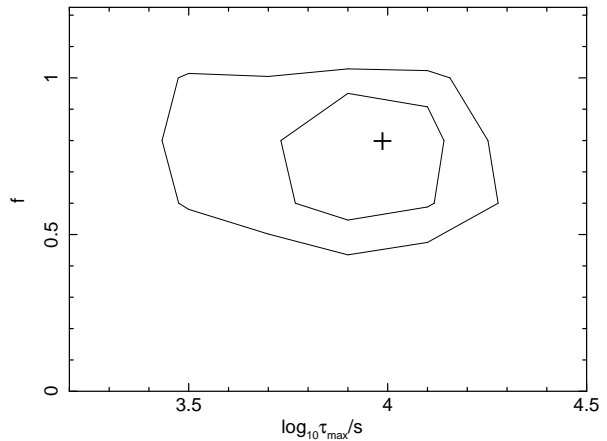}
}}
\caption{
68- and 90-percent confidence intervals on the model
parameters $\log_{10}\tau_{\rm max}$ and $f$ for the simple single-shell 
reverberation model as described
in the text, derived from fitting to the combined 2005-8 dataset.  The cross
marks the best-fit location.
}
\label{fig:contours}
\end{figure}

To place statistical limits on the parameters of the simple thick-shell reverberation model, 
we used the frequency-dependent lag model of equation\,\ref{eqn:lagmodels} within the likelihood
maximisation procedure to find best-fit values of $f$ and $\tau_{\rm max}$, stepping through a
grid of $f, \tau_{\rm max}$ values and evaluating the best-fit likelihood for each grid parameter set. 
The value of the inner radius of reverberation is not well constrained by the data, and this
was fixed at a value $\tau_1=500$\,s.
Confidence levels were assigned by finding intervals in log(likelihood), $\Delta\chi^2 = 2\Delta\ln\mathcal{L}$,
of 2.3 and 4.61 corresponding to 68- and 90-percent confidence intervals.  The best-fit values
for the lags between the 6-10\,keV and 0.3-0.94\,keV bands 
were $f = 0.8^{+0.15}_{-0.25}$, $\log_{10}\tau_{\rm max} = 4.0_{-0.3}^{+0.15}$ 
(i.e. $\tau_{\rm max} = 10\,000_{-5000}^{+4100}$\,s), where quoted errors are 68\,percent confidence intervals,
with a 90-percent confidence range in $\tau_{\rm max}$ of
$2\,500 < \tau_{\rm max} < 20\,000$\,s.
The confidence intervals are shown in Fig.\,\ref{fig:contours}.
As $\tau_{\rm max} = 2r_2/c$, the reverberating region extends to a distance 
$r_2 = 5\,000_{-2500}^{+2050}$\,light-seconds in this model.  We can see from consideration of
Fig.\,\ref{fig:doubleshell} that the double shell model produces more reflection from smaller
radii (in the example shown, up to $r \la 800$\,s) but with additional larger-scale reflection
extending beyond the region where the current data can constrain its extent ($r \ga 10$\,000\,s).
Irrespective of the detailed model,
the continuous nature of the lag spectrum, lags monotonically increasing with time period, implies
there is thick-shell reverberation over a wide range of radii.

Overall, our conclusion is that reverberation naturally reproduces the frequency-dependent lags. 
The reflection fraction $f$ may be directly interpreted
in terms of the global covering factor of the source, $C_g$: in the case where our line of sight to the source
is unobscured, $f=0.8$ implies the remaining sight lines are 80\,percent covered by material if it is
100\,percent reflective, when integrated through the reflecting zone, with higher covering factors
for lower integrated reflection albedo.  However, the inferred global
covering factor is smaller if our line of sight is partially obscured, as this means we could have 
overestimated $f$.
In particular, if the covering
fraction of our line of sight is the same as the global covering factor, and if that material transmits
no light but has an albedo $A$, we expect $f \simeq AC_g/(1-C_g)$
from which we would deduce $C_g = f/(A+f)$.  Thus for $f=0.8$ we deduce $C_g \ga 0.44$ for $A < 1$.

The reverberation time lags we deduce in either the single-shell or more complex models
place a significant component of reflecting material well
within the broad-line region. 
In the case of the single shell model, it extends to 
distances $r_2 \simeq 1.5\times 10^{14}$\,cm, or 600\,r$_g$ 
(where r$_g=GM_{\rm BH}/c^2$) with a 
90\,percent confidence range of 150-1200\,r$_g$ for the nominal black hole mass of 
$M_{\rm BH}=1.7\times 10^6$M$_\odot$.  More complex models may have a component of reflection
extending to larger radii, but this must be counter-balanced by enhanced reflection also
at smaller radii.  The most likely reflection configuration is discussed further in 
section\,\ref{sec:completepicture}.

\subsection{The reflection spectrum}
\label{sec:fspectrum}
\begin{figure}
\resizebox{0.45\textwidth}{!}{
\rotatebox{-90}{
\includegraphics{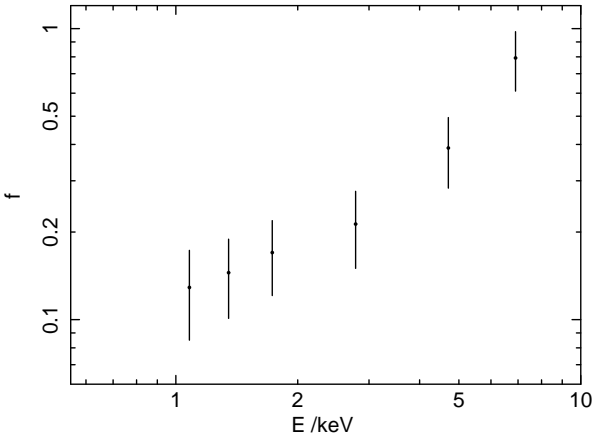}
}}
\caption{
The variation in reverberation model reflection factor, $f$, (the ratio of reflected to directly-received
flux) with the median energy $E$ of each band.
}
\label{fig:fspectrum}
\end{figure}

\citet{mchardy04a} found that the time lags increase with the difference in energy between the bands being
cross-correlated: harder bands are more lagged with respect to softer bands.  This is confirmed by
the analysis here.  In the reverberation model, this observation implies that the fraction $f$ of
reflected light increases with increasing energy.  We can quantify this by fitting the 
simple ``single thick shell'' reverberation
model to the time lags measured in various bands, and for a fixed model value of $\tau_{\rm max}$ we
can deduce $f$ as a function of energy band, again assuming that the softest band has a negligible
reflection component.
This is shown in Fig.\,\ref{fig:fspectrum} 
for the same energy bands used above, for the best-fit value $\tau_{\rm max}=10^4$\,s.
The median photon energy in each band is plotted.

It is striking how similar the energy dependence looks to the spectrum of the offset component in the
PCA analysis, which we have already interpreted as arising from a combination of absorption and
reflection.  We thus have a natural explanation, not only of the frequency dependence of the lags,
but of their energy dependence also.  The fall-off to low energy is a direct consequence of scattering
in a medium whose opacity increases with decreasing energy. It is a generic feature of models of
optically-thick reflection \citep[e.g.][]{magdziarz95a,ross05a}, but is expected to be seen also in
the spectrum of scattered light in a less optically thick medium. A full calculation of Compton
scattering in an extended, ionised, possibly clumpy, medium is required to make a full prediction of
the expected spectrum.  Of particular interest is the strength, line energy and profile of Fe\,K$\alpha$
emission that would be expected.  We saw in section\,\ref{sec:spectralvariability} that there exists a
broadened component of Fe\,K$\alpha$ on eigenvector one with equivalent width $66_{-37}^{+43}$\,eV.
If the reflected light accounts for a fraction $f/(1+f)$ of the total observed light, for 
$f=0.8^{+.15}_{-.25}$ we
find the equivalent width against the reflected component alone\footnote{
One caveat is that 
we need to be careful in interpreting the PCA components in this context: they were derived from spectral
variability on timescales longer than 20\,ks, but in the reverberation signal we are studying variations
on much shorter timescales, and we do not know that the PCA spectral decomposition applies on
these timescales.  If the long-period variations that shape the PCA are caused by partial-covering
variation in the inner regions, some component of the reflection spectrum from the
reverberating region could appear in the offset spectrum rather than being entirely contained
within eigenvector one, and thus the true equivalent width of Fe\,K$\alpha$ on the reflection
component could be higher than estimated here.  
Unfortunately we lack sufficient signal-to-noise to apply the PCA method 
to faster variations while retaining the spectral information.}
would be about $150_{-85}^{+113}$\,eV.  This
is significantly lower than would be expected for reflection from neutral material, where equivalent
widths of order 1\,keV are expected \citep[e.g.][]{george91a}, but in the environment close to the
primary X-ray source we expect gas to be highly ionised, and such equivalent widths are produced
in models of reflection in ionised material, for values of ionisation parameter 
$\xi \ga 100$\,erg\,cm\,s$^{-1}$ \citep[e.g.][]{ross05a}.  Two particular effects that would need to
be taken into account in modelling the expected line emission are the effects of Compton scattering
in a hot medium, which could significantly broaden the line, and the effects of resonant absorption
and resonant line destruction in material where Fe is ionised to higher than Fe\,{\sc xvii} 
(\citealt{matt97a}, see also the review by \citealt{turner09a} and references therein).  

In practice, the reflection spectrum and the dependence of time lag on frequency would be expected
to have a more complex inter-relationship than discussed in this simple model.  Material close to
the ionising source would be expected to have a higher ionisation, and hence be more reflecting
at lower energies, than material further away.  Long period fluctuations would thus have
reflection spectra characterised by lower ionisation than the reflection associated with
short period fluctuations.  The structure and geometry of the reflecting region could also have a
significant effect on the reflected spectrum: for example, a clumpy two-phase medium could simultaneously
produce a ``grey'' reflection spectrum from fully ionised material and a hard reflection spectrum from
lower-ionisation clumps.  Or a highly ionised reflecting region might itself be surrounded by lower
ionisation absorbing material.  Such models could produce a wide range of spectrum hardness and emission
line equivalent widths.  We postpone investigation of more sophisticated models to future work.

\subsection{Coherence and the amplitude of the PSD}
The most noticeable feature of the comparison of the 2008 PSD between hard and soft 
bands is that the high energy bands
show less fractional variability than the soft bands (i.e. the amplitude of the PSD is significantly
higher in the soft bands compared with the hard band).  The coherence in the overall dataset
is high, however, indicating that the same fluctuations are present in soft and hard bands
but that their amplitude in the hard band is suppressed. 
Strangely, though, at the highest frequencies the soft and hard
bands have comparable PSDs, with the hard band variations possibly being even higher than the
soft band, which does not fit easily with the simplest interpretations.  

First, consider the expectation of the reverberation model. In the combined direct plus reflected signal
the power spectrum is expected to be suppressed by a factor $(1+f)^2$ at frequencies
$\omega \ga 1/\tau_{\rm max}$, compared
with the direct signal alone (section\,\ref{sec:reverb}).  Thus if the soft band PSD is dominated
by intrinsic variations, and the hard band PSD is a reverberation-modified version of that, then
we expect the 2008 hard band PSD to be lower than the 2008 soft band PSD
by a factor about $(1.8^{+.15}_{-.25})^2 \simeq 3.2^{+.6}_{-.8}$.
This is consistent with the difference in PSD amplitudes at frequencies $10^{-5} < \nu < 10^{-3}$\,Hz
(Fig.\,\ref{fig:psd}) for $\tau_{\rm max} \ga 10\,000$\,s.

However, this expected PSD amplitude difference is not observed at the
highest frequencies, $\nu > 10^{-3}$\,Hz, in 2008.
Here, the hard band had at least as much variance as the softer band.  
It has previously been suggested that the flatter hard-band PSD is a result
of the mechanism that produces the intrinsic fluctuations \citep[e.g.][]{arevalo06a}, 
which we have not addressed here.  If we believe that the fluctuations we
observe are intrinsic, the flatter PSD could be a 
consequence of that process.  As an example of a possible effect that could cause it,
the highest frequency modes are likely not independent gaussian 
fluctuations, but are more likely the high frequency modes associated with sharp spikes of emission,
so in this picture soft-band spikes might last longer than hard-band spikes.  Without a more specific model
for the emission process it is difficult to make further progress on understanding the PSD.
A future statistical analysis of non-gaussianity in the time series may shed more light on
the high-frequency behaviour.

The coherence of the hard and soft band signals may also aid our understanding.  In 2008 the coherence
was high at all frequencies, with a minimum coherence $\mathcal{R} \simeq 0.5$ at intermediate frequencies,
rising to higher coherence at high, and probably at low, $\omega$ (Fig.\,\ref{fig:coherence}).  
Reverberation is qualitatively expected to show similar behaviour (section\,\ref{sec:reverb}), but
as the maximum observed value of $\omega\tau^\prime \simeq 0.21$ (Table\,\ref{table:lags}), the minimum
coherence expected from reverberation alone is $\mathcal{R} = \cos^2(\omega\tau^\prime) \ga 0.95$.
Reverberation might make some contribution to the dip in $\mathcal{R}$ at intermediate frequencies,
but overall the observed coherence
values in the range $0.5 \la \mathcal{R} \la 0.9$ are more likely due 
either to imperfect intrinsic correlation between source
fluctuations in the soft and hard bands or to some weak additional component of variability such as
might arise from variable obscuration.  Such effects cannot dominate, however, otherwise they would
destroy the reverberation time lag signatures.

Turning to the 2005 low-state data, the PSDs have a different shape from the PSDs in 2008
(section\,\ref{sec:psd}) and show both enhanced soft band variability 
on longer timescales and reduced coherence between
the soft and hard bands.  These two observations are 
indicative of an additional source of variability affecting
the soft band but less-so the hard band in 2005.  Given the spectroscopic evidence
that the central source was obscured in 2005 (section\,\ref{sec:pca}),
a strong candidate for that additional variability
is varying absorption, either changes in column density and ionisation, or variations in
covering fraction.
A varying covering fraction of a clumpy absorber would match the longer timescale 
spectral variability uncovered by the PCA.
In the hard band, some variations arising from the absorber changes would be expected, but
greatly reduced in amplitude because of the lower opacity at higher energies.  As these
variations would be uncorrelated with any intrinsic variations, the soft and hard bands
would become more decorrelated, as observed (Fig.\,\ref{fig:coherence}).

\subsection{The absence of inner-disk reflection signatures}
In Section\,\ref{sec:pca} we saw how there exists a hard component of emission which appears
to be quasi-steady, and essentially represents the low spectral state of the source.
It is this component which, in mean spectra, leads to the existence of a ``red wing'' at
energies below the 6.4\,keV\,Fe\,K$\alpha$ line, and which has previously been widely
interpreted as being highly redshifted reflection from the accretion disk close to or
even inside the last stable orbit of the black hole.  The general lack of variability of this
component is problematic however, as its amplitude should vary with the amplitude of the
illuminating continuum.  To counter this expectation, \citet{fabian03a}, \citet{miniutti03a}
and \citet{miniutti04a} have proposed a model in which the continuum source is compact,
and lies close to the black hole.  The source is hypothesised to move towards and away from
the accretion disk, as it does so light rays follow geodesics in the space time, and with
careful placement it can be arranged that light reflected from the disk and reaching the
distant observer can have an approximately invariant flux, while the flux from the source
itself is low when the source is near the disk and high when it is further away.

The large time lags found here must however come from more distant material, and even though
in NGC\,4051 there {\em is} a component of hard excess that varies with the primary
illumination, the reflecting material must be distant, not from the inner disk.  We cannot
rule out the existence of additional inner disk reflection, with essentially zero time
lag, but given that we already need distant (lagged) reflection there is no justification
for invoking an additional inner disk reflection component.  Furthermore, the hard
spectral component can be modelled quite satisfactorily as being caused by reflection
and/or absorption from intervening material (e.g. in MCG--6-30-15, \citealt{miller08a, miller09a},
see also \citealt{terashima09a} and Lobban et~al. (in preparation) for discussion of NGC\,4051).
In the reverberation signal we find evidence for a high global covering factor of
reflecting components, so all spectral features are accounted for.

\subsection{Towards a complete picture of spectral variability}
\label{sec:completepicture}
In this final discussion section, we attempt to collate the various observational results and interpret them in
terms of the reverberation picture that has emerged in the previous sections.  The aim is to build
a complete model of the source spectral variability that is consistent between the spectral and timing
analyses. 

First, the difference between the 2005 and 2008 spectral states may be explained as being caused by
obscuring material blocking a direct view of the central source, so that the flux we see is dominated
by reflection in 2005.  This explains the strong invariant 6.4\,keV\,Fe\,K$\alpha$ line as well as the 
spectrum continuum shape if the line-emitting material continued
to receive an invariant ionising flux: that is, if most lines of sight other than ours remained open.  
Low, reflection-dominated states of NGC\,4051 have previously been identified by
\citet{guainazzi98a}, \citet{pounds04c} and \citet{terashima09a}.  The obscuration
idea was proposed by \citet{pounds04c} before the {\em Suzaku} data became available, and is confirmed
by the later observations.  \citet{guainazzi98a} and 
\citet{terashima09a} also proposed an invariant reflection component, although
their suggestion that its constant nature is due to light travel time effects is not required if most of
the transition between low and high flux states is caused by obscuring material 
along our line of sight to the central source.
Because we did not have a direct view
of the central source in 2005, any rapid variations would have been
erased by reflection travel time delays, causing the steeper cut-off
to high frequencies in the 2005 PSD compared with the 2008 PSD (Figs\,\ref{fig:obs3comp} \& \ref{fig:psd}).
We do not have any good constraint
on the location of the constant reflection component, however.  The low coherence between soft and hard bands
on timescales $< 5000$\,s and the large offset between the PSD amplitudes for the soft and hard bands in 2005
implies the presence of a significant additional source of variability in the soft band, which we suggest is
variable obscuration, most likely changes in covering factor, during the 2005 observation.

Moving to the 2008 observations, the hard excess increased with the general brightening of the source.
There are two possibilities. If the variable component of hard excess arises from reflection, 
but if the 2005 state was indeed due to obscuration, the postulated variable reflection component
must also have been suppressed in 2005, i.e. it was also partially obscured along our sightline.
This would be consistent with the partial-covering fitting by \citet{terashima09a}.
There is evidence for a moderately broad component of Fe\,K$\alpha$ that varied with the variable continuum
(sections\,\ref{sec:spectralvariability} and \ref{sec:fspectrum}).
The variable hard excess and associated Fe\,K$\alpha$ component would need to come from
a region closer in than the region responsible for the steady, narrower Fe\,K$\alpha$, so that it could
be obscured in this way. 
Alternatively, in 2008 we could still have been looking through some partially-covering absorbing material that
allowed transmission at high energy (as in PDS\,456, \citealt{reeves09a}) and which was suppressed in 2005
because of a higher column density.  In either case, we require the hard excess to be partially obscured
in 2005 and possibly in 2008 also.  Both possibilities require obscuring and reflecting material to exist
over a wide range of radii.  Since 1998, the Fe\,K$\alpha$ strength has been constant to within a 
factor two \citep{terashima09a}: as the emitting region may well be substantially smaller than 10 light
years, the implication is that the intrinsic illuminating continuum likewise has been constant
to within this factor, and that the larger amplitude continuum variability {\em on long timescales} is due to
variations in obscuration along our line of sight.

The time lags and their frequency- and energy-dependence indicate reverberation from reflecting material.
The fraction of reflected light must be high, $f \simeq 0.8$ at $E \ga 6$\,keV, 
indicating a high global covering factor
for this gas.   Model fits of a thick reflecting shell imply the
reverberating zone is within a few light hours of the central source, extending over distances up to
$5\,000_{-2500}^{+2050}$\,light-seconds from the illuminating source.
The spectrum of reflected light required to match the energy dependence of the lags is consistent
with having the hard spectrum generically expected from reflected emission from ionised material.
Obscuring and reflecting material
may, and probably does, exist at larger distances, but we do not expect to see signatures of longer
time lags in the timing analysis presented here, as longer timescales would need to be probed.
We note that reverberation signals are not created by variations in obscuration along the
line of sight alone, because most sight lines would be unaffected.  Thus the primary fluctuations
whose echoes we see in the reflected light must either be intrinsic or, if due to varying obscuration,
must be such that the majority of sight lines are affected.

The implications so far then are that long-timescale variations may be caused by changes in
obscuration along the line of sight (likely to be partial covering changes) but that variations on
timescales of days or less are most likely caused by an intrinsic mechanism.  Because the coherence
between energy bands is high over the observed timescales of days and less in 2008, and because we expect
varying obscuration to be uncorrelated with intrinsic variations, there is not much room for
both varying obscuration and intrinsic variability on these timescales in the more common high states
of the source, such as in 2008.

\section{Conclusions}
We have considered both the spectral variability and the broad-band variability power spectrum of three
{\em Suzaku} X-ray observations of NGC\,4051 spanning three years.  
We find and confirm many features from the spectral analysis and the power spectrum
analysis that have been found by previous authors using independent data and methods, 
both in this AGN and others.
We have related the rapid variability features to the longer-timescale spectral properties, and we 
have shown that a self-consistent picture between these regimes is possible.

The interpretation of the
long-term spectral variability
of NGC\,4051, previously noted by \citet{guainazzi98a}, \citet{pounds04c} and \citet{terashima09a},
as being due to the primary X-ray source disappearing from view, leaving a constant, hard,
reflection-dominated component, is confirmed here by the principal components analysis, and
we reiterate the suggestion that this is due to variable partial covering obscuring the
central source during observed low states,
explaining the almost constant Fe\,K$\alpha$ narrow line flux.
The PSD differs in the 2005 low state from that in the
2008 higher-state, showing features consistent with this explanation.

We have detected the previously-known time lags between hard and soft bands
in the new {\em Suzaku} X-ray light curve of NGC\,4051.
Although previously hypothesised to be in some way related to the 
mechanism that produces X-ray emission, here we conclude that these
time delays, and by inference those in other
AGN also, are well described by reverberation from material up to a few light hours from the central source.
The important signatures are a strong dependence on the frequency of variations being measured 
and on photon energy, both of which have a natural explanation in the reverberation model.
The frequency-dependence is a consequence of observing superimposed direct and reflected components,
and the measured lag-frequency relation
is consistent with reverberation from a thick shell of material, although
the precise geometry is not yet well constrained by the data.  The measured energy-dependence implies a
hard reflection spectrum that is consistent with that expected from ionised material
and which is also similar to the hard component that is seen in principal components analysis of the
spectral variations.  
To obtain sufficient reflection, the covering factor of reflecting material
must be high, greater than about 0.44, as recently inferred from spectral modelling of other
AGN \citep[e.g.][]{miller09a}.  The existence of a maximum observed time lag at low frequencies
and possible negative time lags at high frequencies, as observed in Ark\,564 \citep{arevalo06a, mchardy07a}
are expected in the reverberation model.

The identification of the signal as being reverberation
confirms the idea that NGC\,4051 and similar AGN are surrounded by substantial
covering factors of optically-thick, or nearly so, material within a few light-hours of the 
central source (far inside both the optical broad-line region and the dust sublimation radius).
Although this AGN has a spectral hard-band excess very similar to that seen in other type\,I
AGN, there is no requirement for any relativistically-redshifted
reflection from the inner accretion disk, which would not generate the observed time lags.
The reverberating material extends over distances up to a few hundred
gravitational radii from the illuminating source.
Given the spectral and timing similarities with other type\,I AGN, we suggest that this source
structure is a common feature of the type\,I population.

\vspace*{1ex}
\noindent
{\bf Acknowledgments}.
This research has made use of data obtained from the {\em Suzaku} satellite, 
a collaborative mission between the space agencies of Japan (JAXA) and the U.S.A. (NASA). 
TJT acknowledges NASA grants NNX09AO92G and GO9-0123X.

\bibliographystyle{mn2e}
\bibliography{xrayreview_aug09}

\label{lastpage}

\end{document}